# Microscopic insight on the pump-probe relaxation dynamics of superconductors: Model study of MgB2 relaxation within nonlinear response theory

Pavol Baňacký[*] and Vojtech Szöcs[♦]

Chemical Physics division, Institute of Chemistry, Faculty of Natural Sciences, Comenius University, Mlynska dolina CH2, 84215 Bratislava, Slovakia

Here we present a quantum-statistical formulation of relaxation dynamics of superconductors related to pump-probe (PP) spectroscopy. The method is based on the perturbation expansion of the non-equilibrium density matrix for calculation of multi-time correlation functions, and the corresponding response function. As a model for our study, the high-temperature superconductor $MgB_2$ has been selected. Knowledge of the electronic structure and of the corresponding Eliashberg function of electron-phonon (EP) interactions enabled us to distinguish non-equilibrium processes in a sequence of interactions with laser pulses on a microscopic level. Temperature-dependent relaxation dynamics as a function of delay time between the pump and probe pulses have been derived. We have shown that an abrupt increase of the relaxation time at $T_c$ is the direct consequence of sudden changes in the character of EP coupling in transition from an adiabatic to a stabilized superconducting anti-adiabatic state, as it predicts the anti-adiabatic theory of electron-vibration interactions. The BCS model, which is based on adiabatic character of EP coupling, is basically unable to reflect the enormous sudden increase of the relaxation time at $T_c$. Differences in the optical pump-optical probe and the optical pump-terahertz probe settings of PP relaxation dynamics are discussed within diagrammatic perturbation theory.

## 1. Introduction

Pump-probe (PP) spectroscopy, a method of nonlinear optics, is an important experimental tool in the study of ultrafast relaxation dynamics of excited states in condensed matter systems [1,2]. The interpretation of experimental results relies on various types of phenomenological equations of motion for the relaxation of "hot electrons" and/or high-energy phonons in coupling to bosonic degrees of freedom [3-7]. The assumption of quasi-equilibrium, which is often behind these methods, is discussed in [8], where the authors present an analytic solution of excited state relaxation dynamics within the Boltzmann kinetic equation, the method of which deals with electrons and phonons out of equilibrium.

In general, however, phenomenological treatments not only ignore nonlinear aspects of processes running in PP experiments, but ignore also the microscopic character of non-equilibrium states – coherences and/or populations, which can evolve in the studied system over a sequence of different time periods via the interaction of matter with the incident fields. In these circumstances, mainly in the case of superconductors, it is difficult to disclose the relationship between the derived relaxation parameters, i.e. relaxation times, or electron-phonon (EP) coupling constants (λ), and the microscopic mechanism of the superconducting state transition. The theory of high-temperature superconductors (e.g. cuprates, pnictides) is far from being settled, and the character of the bosonic field, which in interaction with an electronic subsystem triggers transition into a superconducting state, remains unclear and open to discussion. In these circumstances, the interpretation of different relaxation times for time scales fs/sub-ps, ps and 100s-ps also remains controversial; these times are extracted by fitting experimental data to 1–3 single exponential decay functions. There is general agreement, however, that EP interactions are responsible for transition into a superconducting state in classical low-temperature superconductors, as well as in the high-temperature superconductor $MgB_2$.

---

[*] corresponding author: banacky@fns.uniba.sk ; [♦] v.szocs@gmail.com

No matter what supposed character of the interacting bosonic field, time-resolved PP experimental results for cuprates, pnictides, $MgB_2$ and simple metals have disclosed the effect that is characteristic and common to all types of these superconductors, and can be detected in studies of the temperature dependence of relaxation dynamics. In particular, it is an abrupt increase of the relaxation time $\tau_r$ at critical temperature $T_c$ when in the cooling (or heating) process, the superconducting gap of a particular superconductor is open (or closed). Hence, it is justified to expect that this effect is closely related to the microscopic mechanism of superconducting state transition. For BCS superconductors, this effect was formulated by Tinkham [9] in the form of a quasi-divergence of the relaxation time at $T_c$, $\tau_r \propto \Delta(0)/\Delta(T_c)$.

As mentioned above, an abrupt increase of $\tau_r$ at $T_c$ is a general attribute detected for all classes of superconductors; and an experimentally extracted time scale where this effect is detected can be used as an important parameter in the assessment of the microscopic mechanism of superconducting state transition. In this respect, the relaxation dynamics of $MgB_2$ (as a thin film on different substrates) derived from optical pump-optical probe (OPOP) and optical pump-terahertz probe (OPTP) experiments, seem to be rather confusing. In OPOP experiments [10] (where the relative change of reflectance is detected), the relaxation dynamics at high temperatures (T ~ 70–300 K) are dominated by single-exponential fast sub-ps decay, with $\tau_{r1} \approx 160$ fs. Below 70 K, besides $\tau_{r1}$ dynamics, the next relaxation process $\tau_{r2}$ on the ps time scale is identified. The $\tau_{r2}$ (~ 1–3.5 ps) dynamics are strongly T-dependent, and at critical temperature $T_c$ (~ 35 K) exhibit a characteristic quasi-divergence. Third, the very slow process is present only below $T_c$, with $\tau_{r3}$ ~ 400 ps. This process is temperature-independent, but is strongly substrate-dependent. The different character of dynamics is presented by OPTP experiments [11], where a change in conductivity (proportional to transmittance) is detected. Though comparable to OPOP experiments [10], three time scales ($\tau_{r1}$ ~ sub-ps, $\tau_{r2}$ ~ few-ps and $\tau_{r3}$ ~ 100s-ps) are identified also in this case, while the difference between these results is substantial. A sudden increase of the relaxation time at $T_c$ is not detected in OPTP experiments [11] for $\tau_{r2}$ dynamics but for the slow $\tau_{r3}$ ~ 400 ps process. In the same paper [11], the authors present results of OPOP measurements, and a sudden increase of relaxation time at $T_c$ is detected again for the slow $\tau_{r3}$ ~ 400 ps process, as in the case of OPTP.

We do not intend to speculate on the reason for difference in the two reported [10,11] OPOP experiments, since it should be related to the quality of the prepared thin films and/or the character of the substrates. In any case, however, the difference in the relaxation dynamics of the order of $10^2$ between OPOP [10] and OPTP [11] for a sudden T-dependent effect at $T_c$, which should be related to the microscopic mechanism of superconductivity, strongly indicates that over the time period between the pump and probe pulses, different microscopic processes might be detected.

On the phenomenological level of the dynamics description, the basic character of the running processes over the time periods when matter interacts with a sequence of incident fields, and the connection of these processes to the microscopic mechanism of relaxation dynamics and to the recovery of a superconducting state, remains unclear. In these circumstances, it seems meaningful to attempt to investigate the relaxation dynamics of superconductors in PP experiments from a microscopic standpoint. Such a treatment has already been developed for relaxation dynamics of semiconductors [12], but comparable microscopic theory for the relaxation dynamics of superconductors is, to our knowledge, absent.

It should be mentioned, however, that in recent years several papers devoted to a microscopic description of the interaction of superconductors with external electric fields (THz or optical laser pulse) have been published [13-18]. All these studies are based on non-equilibrium density matrix methods, but they are basically restricted to a linear optical process. The focus is placed mainly on the time evolution of BCS-order parameter (i.e., the BCS gap expressed through t-dependent Bogoliubov quasi-particle densities) after an interaction of a system with an incident pump field that depends on the pump pulse frequency, duration and intensity. The results display interesting aspects and conditions for quantum oscillations and the generation of coherent phonons. Nevertheless, the aspects of the interaction of matter with a sequence of incident fields, which is the basic attribute of any PP experiment, is beyond the scope of these studies.

Here we present a quantum-statistical formulation of third-order polarization $P^{(3)}(t)$, which is induced in a sample by a sequence of incident external fields, and which serve as the source of an emitted radiation field detected in PP experiments. This treatment is based on the perturbation expansion of the non-equilibrium density matrix for calculation of multi-time correlation functions, and the corresponding response function at finite temperatures. As a model for our study, the high-temperature superconductor $MgB_2$ has been selected. This material is simple enough for theoretical study, and with its two-gap character it represents the aspects of superconductivity in full complexity. Moreover, the crucial role of EP interactions in transition into a superconducting state is generally accepted and the Eliashberg function $\alpha^2 F(\omega)$, which represents the strength of these interactions, is for $MgB_2$ well established, along with the detailed decomposition on contributions of particular phonon branches [19]. It enables us to treat the dissipative part of excited state dynamics on a quantum mechanical level in a realistic way.

The present study is focused towards numerical modeling of the temperature dependence of relaxation time, $\tau_r(T)$, which is for a particular temperature, extracted from the signal evolution $S(t_D)$, calculated as a time-integrated function of third order polarization $P^{(3)}(t, t_D)$, which explicitly depends on the delay time between the pump and probe pulse, $t_D$. Due to theoretical and numerical complexity, in this paper we only present the results for OPOP relaxation dynamics calculated for two models of superconductivity, which are both based on the EP mechanism in the superconducting state transition. The first model (denoted as BCS) is related to adiabatic character of EP interactions, as it assumes the BCS theory [20-23,39], while the second model respects the character of EP interactions for the adiabatic to anti-adiabatic state transition, with the breakdown of the Born Oppenheimer approximation at $T_c$, as it follows from anti-adiabatic theory (AAT) of the superconducting state transition [21-25]. The numerical solution to the relaxation dynamics yields, for the AAT model, an abrupt increase of the relaxation time on a few-ps time scale at two distinct temperatures, $T_{c\sigma}$ and $T_{c\pi}$ ($T_{c\sigma} > T_{c\pi}$), which correspond to critical temperatures for the opening of the larger gap $\Delta_\sigma$ in the σ band, and a smaller gap $\Delta_\pi$ in the π band.

Within the BCS model, changes of the relaxation time near $T_{c\sigma}$ and $T_{c\pi}$ are less pronounced, in spite of the fact that the temperature dependence of $\Delta_\sigma(T)$ and $\Delta_\pi(T)$ is treated for both models in the same way. Whereas the relaxation dynamics at high temperatures (T > $T_{c\sigma}$) are identical for these models, the different temperature dependence of the relaxation time below $T_{c\sigma}$ is related to the different treatment of σ-σ and σ-π interactions within the BCS and AAT models in the superconducting state. This suggests that, for the BCS adiabatic state, a single fact of superconducting gap opening and its temperature dependence is a weak physical effect to induce quasi-divergence in the temperature dependence of excited state relaxation dynamics. The AAT results indicate that the abrupt change in relaxation time at $T_c$ is strongly connected to a sudden change in character of EP interactions crossing from an adiabatic into a stabilized antiadiabatic superconducting state, with broken local symmetry, and with gap(s) opening in the single particle spectrum.

The paper is organized as follows: In section 2, with reference to details in Appendix A, is a short sketch of the microscopic background of PP spectroscopy. The models of $MgB_2$ for excited state relaxation dynamics are specified in section 3, along with a specification of EP coupling and of derived forms of electronic and vibration parts of relevant response functions. The results of numerical calculations of relaxation dynamics for our models obtained through the Meier-Tannor decomposition scheme for spectral density [26,27], which is derived from a particular Eliashberg function [19], and from Padé spectral decomposition of a Bose function [28] for calculating the line-shape function across a wide range of temperatures (5–300 K), are presented in section 4. With respect to published experimental OPOP and OPTP results, the theoretical outputs are discussed on the basis of time-dependent diagrammatic perturbation theory in section 5. All necessary details of applied theoretical and numerical methods used are presented in Appendices A–E.

**2. Density matrix treatment in modeling of PP spectroscopy**

The theoretical basis of our study closely follows a formulation of PP spectroscopy as a special case of four-wave mixing (FWM) spectroscopy [29].

The crucial quantity for the description of coherent nonlinear spectroscopy is an optical polarization P(t), i.e. a collective dipole moment induced in a sample by an incident optical field E(r,t) as a coherently oscillating charge distribution. In theoretical modeling, polarization can be expressed as the mean value of the dipole moment $\langle \mu(t) \rangle$, and can be calculated through a density matrix that obeys the Liouville equation. Relevant for FWM spectroscopy is 3$^{rd}$-order polarization, $P^{(3)}(t) = Tr(\mu \rho^{(3)}(t))$, the source of emitted field detected in PP experiments as a signal. Polarization $P^{(3)}(t)$ can be evaluated through the 3$^{rd}$-order response function $S^{(3)}$, which is represented [29] by a set of four independent multipoint dipole moment correlation functions $\{R_i(t)\}$. Within a multimode quantum Brownian oscillator model, these functions can be factorized on the contributions of electronic transition dipole moment correlations and vibration correlation functions, which reflect electron-vibration interactions in the form of the corresponding line-shape function *g(t)*.

For the sake of brevity, the necessary theoretical background of this formulation is presented in Appendix A.

## 3. Model of MgB$_2$ for the study of excited state relaxation dynamics
### 3.1. Model of electronic structure

The anti-adiabatic model of superconducting state transition (AAT) follows from the explicit dependence of electronic structure on instantaneous nuclear coordinates $\{Q\}$ and momenta $\{P\}$, i.e. on nuclear dynamics. It is a problem beyond the Born-Oppenheimer approximation (BOA), and it was studied [21-24] in details. We present here the final form of corrections to zero and one-particle electronic energy terms, which are the consequence of non-adiabatic electron-vibration coupling, and are relevant for our model. Corrections are, with respect to the corresponding energy terms, calculated for a system within the crude adiabatic approximation, i.e. the standard solution of electronic structure calculation at a frozen-clumped nuclear configuration within the BOA.

Zero-particle correction, i.e. correction to the electronic ground-state energy due to electronic system interaction with the phonon mode $r$ is,

$$\Delta E^0 = \sum_{A,I,r} \left|u_{AI}^r\right|^2 \frac{\hbar \omega_r}{\left(\varepsilon_A^0 - \varepsilon_I^0\right)^2 - (\hbar \omega_r)^2} = 2 \sum_{\varphi_{Rk'}}^{bands} \sum_{\varphi_{Sk}}^{bands} \int_0^{\varepsilon_{k',max}^0} n_{\varepsilon_{k'}^0}\left(1 - f_{\varepsilon_{k'}^0}\right) d\varepsilon_{k'}^0 \int_{\varepsilon_{k,min}^0}^{\varepsilon_{k,max}^0} \left|u_{k-k'}^r\right|^2 n_{\varepsilon_k^0} f_{\varepsilon_k^0} \frac{\hbar \omega_r}{\left(\varepsilon_k^0 - \varepsilon_{k'}^0\right)^2 - (\hbar \omega_r)^2} d\varepsilon_k^0. \quad (1)$$

Transcription of discrete-level representation to the reciprocal space representation of periodic solids is based on following correspondence: {I}-occupied states below Fermi level (FL) → $\{k,\sigma\}$ with occupation number $f_k$, {A}- unoccupied states above FL → $\{k',\sigma'\}$ with occupation number $(1 - f_{k'})$, $\varepsilon_I^0 \to \varepsilon_k^0$, $\varepsilon_A^0 \to \varepsilon_{k'}^0$, normal mode vibration r→ q, and electron-vibration coupling $\left|u_{AI}^r\right| \to \left|u_{k',k}^q\right| = \left|u^q\right| = \left|u^{k-k'}\right|$. The Fermi-Dirac populations $f_{\varepsilon_k^0}, f_{\varepsilon_{k'}^0}$ introduce into Equation (1) temperature dependence. The term $u_{k-k'}^q$ stands for the EP coupling matrix element of bilinear form and, $n_{\varepsilon k}$, $n_{\varepsilon k'}$ are density of states (DOS) of interacting bands at $\varepsilon_{k'}^0$ and $\varepsilon_k^0$. It is evident that for adiabatic systems, such as metals with large Fermi energy ($E_F$) and with adiabatic ratio $\omega/E_F \ll 1$, the $\Delta E^0$ correction is positive and is negligible. Only for systems undergoing transition into an anti-adiabatic state due to EP interactions when $\omega/E_F > 1$, i.e. in the case of instability of adiabatic electronic structure with respect to some vibration motion, the correction $\Delta E^0$ is negative, its absolute value depending on the magnitudes of $u_{k-k'}^r$ and DOS $n_{\varepsilon k}, n_{\varepsilon k'}$ for vibration displacement $R_d$ in a mode that induces instability that is manifest as FL crossing by some fluctuating electronic band. In the moment when the analytical critical point (ACP, e.g. a maximum) of the fluctuating band $\varphi$ approaches FL, the system not only undergoes transition into an anti-adiabatic state (for relevant electronic energy spectrum $\Delta\varepsilon^0$ now holds, $\Delta\varepsilon^0 < \omega$ and $\omega/E_F > 1$), but the DOS, $n_\varphi(E_F) = \left(\partial \varepsilon_\varphi^0 / \partial k\right)_{E_F}^{-1}$, of the fluctuating $\varphi$-band is considerably increased near FL. It is a situation that invokes a dramatic decrease of effective electron velocity, i.e. the possibility of van Hove singularity formation in this region of DOS.

It should be noted that on a crude adiabatic level, nuclear displacement out-of equilibrium ($R_d$) is always connected with an increase of total electronic energy $\Delta E_d$,

$$\Delta E_d(R_d) = E_0^{te}(R_d) - E_0^{te}(R_{eq}) > 0. \tag{2}$$

The stabilization (condensation) energy $E_{cond}^0$ at the transition from an adiabatic into an anti-adiabatic state is,

$$E_{cond}^0 = \Delta E_d(R_d) - \left|\Delta E_{(na)}^0(R_d)\right|. \tag{3}$$

For the anti-adiabatic state, ground-state energy (1) is negative, $\Delta E_{(na)}^0(R_d) < 0$. Then, if the inequality $\left|\Delta E_{(na)}^0(R_d)\right| > \Delta E_d(R_d)$ holds, the system is stabilized in distorted geometry $R_d$, i.e. local symmetry breaking occurs.

Corrections to the one-particle term [23a,b] due to the dependence of electronic structure on nuclear coordinates and momenta are, for states of band $\varphi_P$ in interaction with states in band $\varphi_{R\,(S)}$ mediated by phonon mode $r$:

$$\Delta\varepsilon(Pk') = \sum_{Rk'_1 > k_F} |u^q|^2 (1 - f_{\varepsilon^0 k'_1}) \frac{\hbar\omega_{k'-k'_1}^q}{\left(\varepsilon_{k'}^0 - \varepsilon_{k'_1}^0\right)^2 - \left(\hbar\omega_{k'-k'_1}^r\right)^2} - \sum_{Sk < k_F} |u^{k-k'}|^2 f_{\varepsilon^0 k} \frac{\hbar\omega_{k'-k'_1}^q}{\left(\varepsilon_{k'}^0 - \varepsilon_k^0\right)^2 - \left(\hbar\omega_{k'-k'_1}^q\right)^2} \tag{4a}$$

for $k' > k_F$, and

$$\Delta\varepsilon(Pk) = \sum_{Rk'_1 > k_F} |u^q|^2 (1 - f_{\varepsilon^0 k'_1}) \frac{\hbar\omega_{k-k'_1}^q}{\left(\varepsilon_k^0 - \varepsilon_{k'_1}^0\right)^2 - \left(\hbar\omega_{k-k'_1}^q\right)^2} - \sum_{Sk_1 < k_F} |u^q|^2 f_{\varepsilon^0 k} \frac{\hbar\omega_{k-k'_1}^q}{\left(\varepsilon_k^0 - \varepsilon_{k_{1_1}}^0\right)^2 - \left(\hbar\omega_{k-k'_1}^q\right)^2} \tag{4b}$$

for $k \le k_F$.

The replacement of the discrete summation by integration, $\sum_k \ldots \to \int n(\varepsilon_k)$, introduces DOS $n(\varepsilon_k)$ into Equations (4a,b). This replacement is of crucial importance in relation to a fluctuating band. For the corrected DOS $n(\varepsilon_k)$, which is the consequence of the shift $\Delta\varepsilon_k$ of orbital energies, the following relationship can be derived:

$$n(\varepsilon_k) = \left|1 + \left(\partial(\Delta\varepsilon_k)/\partial\varepsilon_k^0\right)\right|^{-1} n^0(\varepsilon_k^0), \tag{5a}$$

where the term $n^0(\varepsilon_k^0)$ stands for the uncorrected DOS of the original adiabatic state of the particular band,

$$n^0(\varepsilon_k^0) = \left|\partial\varepsilon_k^0/\partial k\right|^{-1}. \tag{5b}$$

Close to the **k**-point where the original band that interacts with the fluctuating band intersects FL, the occupied states near FL are shifted downward below FL, while the unoccupied states are shifted upward above FL. This means that transition into an anti-adiabatic state induces the formation of the **k**-dependent gap $\Delta_k(T)$ in quasi-continuum of an adiabatic one-electron spectrum, which has been derived [23b] in the form,

$$\Delta(T) = \Delta(0) tgh(\Delta(T)/4k_B T). \tag{6}$$

The gap opening is related to the shift $\Delta\varepsilon_{Pk}$, according to Equations (4a,b), of the original adiabatic orbital energies $\varepsilon_{Pk}^0$, $\varepsilon_{Pk} = \varepsilon_{Pk}^0 + \Delta\varepsilon_{Pk}$, and to the **k**-dependent change of the DOS of particular band(s) at the FL – spectral weight transfer. If there is a phonon coupling with two different bands that intersect the FL in a given direction of the reciprocal lattice at points $k_1$ and $k_2$, then two different gaps in the one-electron spectrum near to these points are opened. This is the case of $MgB_2$, where $\sigma$ and $\pi$ gaps are opened along the Γ-K (Γ-M) direction due to $\sigma-\sigma$ and $\sigma-\pi$ coupling mediated by $E_{2g}$ phonon mode.[25]

With respect to $\Delta(0)$, Equation (6) yields a simple approximate relationship for the critical temperature $T_c$ of the adiabatic - anti-adiabatic state transition,

$$T_c = \Delta(0)/4k_B . \qquad (7)$$

In the stabilized anti-adiabatic state, the ground state total electronic energy of the solid state system is geometrically degenerated. Distorted nuclear structure, related to the couple of nuclei in the phonon mode $r$ that induce the transition into an anti-adiabatic state, has a fluxional character. There exist an infinite number of different and distorted configurations of the couple of nuclei involved in the phonon mode $r$ and, all these configurations, due to translation symmetry of the lattice, have the same ground state energy. Supercarriers are itinerant bipolarons which move in the form of polarized inter-site charge density distribution in real space. Their motion is dissipation-less until system remains in stabilized anti-adiabatic state. It follows directly from the theory.

The effective EP interactions that cover an influence of nuclear dynamics on electronic structure [23a,b, 24], have the form (the expression is for boson vacuum):

$$\Delta H'_{\langle e-p0 \rangle}(dg) = \sum_{k,q(r)} |u^q|^2 \{(\varepsilon_k^0 - \varepsilon_{k-q}^0)/[(\varepsilon_k^0 - \varepsilon_{k-q}^0)^2 - (\hbar\omega_{q(r)})^2]\} N[a_k^+ a_k] . \qquad (8)$$

For an extreme nonadiabatic limit, i.e. for the anti-adiabatic state, $\hbar\omega_{q(r)}/|\varepsilon_k^0 - \varepsilon_{k-q}^0| \to \infty$, the contribution of the phonon mode $r$ that induces fluctuation of some band is:

$$\Delta H'_{\langle e-p0 \rangle_r}(dg)_{na} \to 0 . \qquad (9a)$$

This means that, for electrons that satisfy the anti-adiabatic condition with respect to the interacting phonon mode $r$, in the particular direction of the reciprocal lattice where the gap in the one-electron spectrum has been opened, the electron (nonadiabatic polaron)-renormalized phonon interaction energy equals zero. Due to thermal excitations, at some critical temperature $T_c$ the nonadiabatic correction to the electronic ground state energy, Equation (1), $\Delta E_{(na)}^0(R_d)$, becomes smaller than $\Delta E_d(R_d)$ and the sudden transition from an anti-adiabatic state into an adiabatic state is induced. For $T > T_c$, the system is stabilized in equilibrium geometry $R_{eq}$, which is identical to the $R_{eq}$ that corresponds to high-symmetry structure on a crude-adiabatic level. The system is now in the adiabatic state. This means that for $T > T_c$, the band structure at Fermi level must be represented by a quasi-continuum of occupied and unoccupied states without the gap in the one-electron spectrum that has been opened at 0 K.

In the adiabatic state, $\hbar\omega_q / |\varepsilon_k^0 - \varepsilon_{k-q}^0| \to 0$, properties of electrons are different. From Equation (8) for EP interaction energy now follows:

$$\Delta H'_{ep}(dg)_{ad} \to \sum_{qk} |u^q|^2 \frac{1}{(\varepsilon_k^0 - \varepsilon_{k-q}^0)} \; . \tag{9b}$$

Equation (9b) basically represents the energy of adiabatic polarons.

Anti-adiabatic theory was applied in the study of different classes of superconductors, including HTS cuprates and A15 compounds [21,22,25,31,32]. It was shown that the electronic and thermodynamic properties of the system in a stabilized anti-adiabatic state correspond to the properties of that particular system in a superconducting state. Recently, this theory has been applied for the prediction of superconductivity in new boron-based nanotubular materials [33,34]. Theoretical study of anti-adiabatic state transition in MgB$_2$ has been published [25], with the calculated parameters being in good agreement with experiments [35,36].

The AAT model for the study of the excited state relaxation dynamics is formulated on the basis on these results, and the main features are schematically displayed in Figure 1.

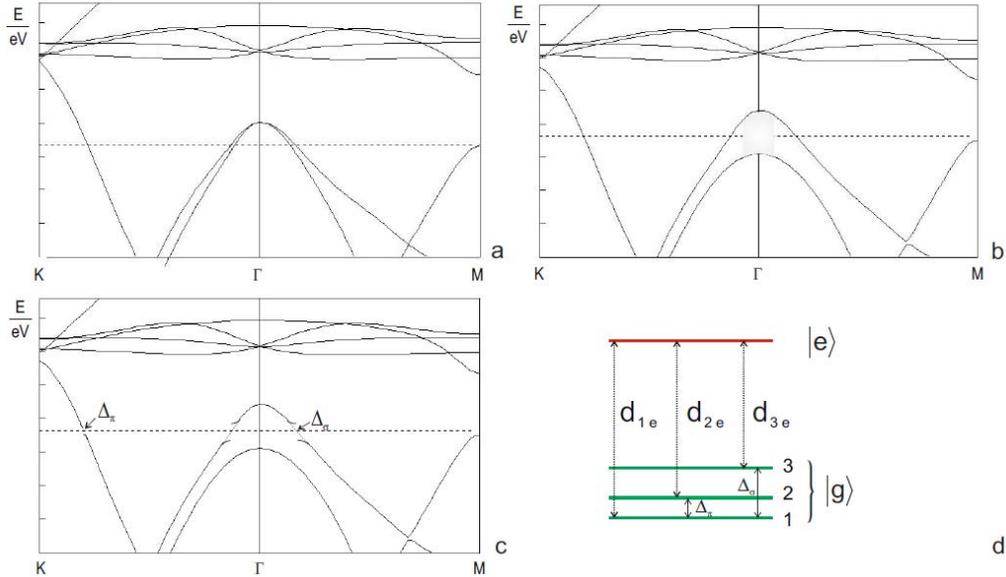

**Figure 1** Adiabatic band structure of MgB$_2$ for (a) equilibrium geometry and (b) for distorted structure with B-atom displacements in B-B stretching vibration, which induces splitting of σ-bands degeneracy, fluctuation of a σ-band across FL (shadow area in Γ point) and breakdown of the BOA. Stabilized anti-adiabatic state (c) with opened gaps $\Delta_\sigma(0)$ =14 meV, $\Delta_\pi(0)$ = 5.6 meV in the σ and π-bands. A simple model of MgB$_2$ electronic structure on the global energy scale for study of relaxation dynamics is shown in panel (d).

The electronic band structure in equilibrium geometry (adiabatic-clumped nuclear Hamiltonian), shown in Figure 1a, is characterized by the degeneracy of two σ-bands in Γ point, forming the hole-pocket above FL. The energy difference between the maximum of σ-bands and chemical potential is ~ 0.5 eV (Fermi energy $E_F$). The π-band intersects FL in the Γ-K direction approaching anti-bonding states, the region of optical excitations, well above FL at K-point. The energy scale of optical phonon modes of MgB$_2$ is in the range of ω ~ 40–90 meV [37], which means that MgB$_2$ in frozen equilibrium nuclear geometry can be treated as an adiabatic system, i.e. $\omega/E_F < 1$. The situation is changed, however, when the electronic system is coupled to vibration motions. In particular, coupling to $E_{2g}$ mode (*ab*-plane B-B stretching vibration) induces splitting of σ-bands degeneracy [25,37-39] and, for B-atom displacements ~ 0.032 Å/B out of equilibrium in the B-B bond, the top of one of the σ-bands is shifted below FL (Figure 1b). In terms of vibration motion, this means fluctuation of a σ-band maximum across FL. At the point when the top of the band approaches FL on an energy distance smaller than $\pm \omega_{E_{2g}}$ ($\approx \pm 70$ meV), the Fermi energy decreases correspondingly, and $\omega/E_F > 1$. In these circumstances, not only is Migdal-Eliashberg (ME) theory invalid (ME holds true only if $E_F$ is much larger than any other energy scale of the system) [38,39], but BOA is suddenly broken. Anti-adiabatic theory [21-23] reflects this possibility and the calculated [25] correction to the ground state energy, see Equation (1), is in this case $\Delta E^0_{(na)}(R_d) =$ -57 meV. The σ-σ interactions add to this value by 48 meV, and the contribution of the σ-π interactions is 9 meV. The increase of total electronic energy on an adiabatic level due to nuclear distortion is $\Delta E_d =$ +12 meV, and the net effect is of stabilization of the system in an anti-adiabatic state in distorted geometry (i.e. local symmetry breaking is induced with $R_d \approx$ 0.032 Å /B in B-B bond), with condensation energy, given by Equation (3), of $E^0_{cond} =$ -35 meV. Calculated corrections [25] to one-particle terms (Equations 4a,b) yield the opening of two gaps in one-particle spectrum: $\Delta_\sigma(0) =$ 15.2 meV in σ-band and $\Delta_\pi(0) =$ 4.4 meV in π-band , with T-dependence according to Equation (6). The resulting form of band structure at 0 K is schematically drawn in Figure 1c. The gaps are closed at a particular $T_c$, according to Equation (7). Above the $T_{c\sigma}$, $E^0_{cond} \geq 0$ and the system obeys the adiabatic condition with nuclear geometry, as it corresponds to the equilibrium structure on an adiabatic level (Figure 1a). It should be stressed that the remaining optical phonon modes in MgB$_2$ (i.e. $E_{1u}$, $A_{2u}$ and $B_{1g}$) do not induce electronic structure instability.

Based on these results, we set up a simplified model of electronic structure on global energy scale for relaxation dynamics study in MgB$_2$ (Figure 1d). The electronic ground state of the system, $|g\rangle$, is represented by three energy levels: $\{\varepsilon_1, \varepsilon_2, \varepsilon_3\}$. At the temperature 0 K, for the opened gaps in the one-particle spectrum holds, $\Delta_\sigma(0) = (\varepsilon_3 - \varepsilon_1)$ and $\Delta_\pi(0) = (\varepsilon_2 - \varepsilon_1)$ with level populations $p_1(0) = 1$, $p_2(0) = 0$, $p_3(0) = 0$. At temperatures other than absolute zero, $T \neq 0$, the magnitude of the gaps evolves according to Equation (6), and the population of levels obeys a Boltzmann distribution. The particular gap is closed at the corresponding $T_c$ according to Equation (7), i.e. $\Delta_\pi$ at $T_{c\pi}$ and $\Delta_\sigma$ at $T_{c\sigma}$. For numerical modeling purposes, we have used experimental gap values $\Delta_\sigma(0)$ =14 meV, and $\Delta_\pi(0) =$ 5.6 meV. The excited state of the system, $|e\rangle$, is represented by a single energy level $\varepsilon_e$. The energy difference $(\varepsilon_e - \varepsilon_1)$ represents vertical optical excitation energy (electronic excitation gap-$\Delta V(Q_{eq})$; Equation (A18) in Appendix A) and for resonant conditions in PP experiments, it corresponds to the energy of a pump pulse, $\hbar\omega_p = (\varepsilon_e - \varepsilon_1) =$ 1.5 eV. It is obvious that for the temperature range $0 \leq T < T_{c\pi}$, the system is represented by a 4-level model {1,2,3,e}, while in the temperature range $T_{c\pi} < T \leq T_{c\sigma}$, the system corresponds to a 3-level model {1,3,e}, and for $T > T_{c\sigma}$, the system is transformed into an effective 2-level model {1,e}. In Figure 1d, the non-zero electronic transition dipole moments are depicted as $d_{ie}$ {i=1,2,3}.

For the BCS model of the present study, the model of global electronic energy levels (Figure 1d) remains formally the same, just that the numerical factor of 4 for T-dependence of the single particle gap in Equations (6) and (7) is replaced by the BCS-value of 3.53. In this case, understanding of the single-particle gap follows the BCS notion, $\Delta=2\Delta_{BCS}$. The difference between the AAT and BCS models is behind the treatment of EP interactions, as specified below.

### 3.2. Model of electron-phonon interactions

The crucial aspect for studying the relaxation process is modeling the dissipation interaction of electronic subsystems with an environment, represented by EP interactions with pertinent phonon modes. The Eliashberg function $\alpha^2 F(\omega)$ is the quantity that characterizes this effect through the second moment of this function,

$$\lambda\langle\omega^2\rangle = 2\int_0^\infty \alpha^2 F(\omega)\omega\,d\omega. \qquad (10)$$

The dimensionless EP coupling constant $\lambda$ is calculated as an integral norm, which represents a key parameter of BCS-like theories of superconductivity.

The Eliashberg function of $MgB_2$ is well documented theoretically as well experimentally, and has been decomposed [19] within multi-band Eliashberg theory on the particular contributions of the participating phonon branches; $\alpha^2 F_{\sigma\sigma}(\omega)$, $\alpha^2 F_{\sigma\pi}(\omega)$, $\alpha^2 F_{\pi\pi}(\omega)$, $\alpha^2 F_{\pi\sigma}(\omega)$. This form of Eliashberg function and its decomposition serves as an input for modeling relaxation dynamics in the present study.

We performed numerical conversions of published types of dimensionless Eliashberg functions [19], and have constructed new and simplified model functions, which best reproduce the published data and keep the corresponding norms $\{\lambda_{\beta\gamma}\}$. The numerical fit of the model $\alpha^2 F_{\beta\gamma}(\omega)$ functions into a sum of a proper number of Lorenzian components shows that, for every type of coupling, some characteristic frequencies $\Omega_k$ are substantial. As an example, we present here the decomposition of $\lambda_{\pi\pi}, \lambda_{\sigma\pi}$ and $\lambda_{\sigma\sigma}$ types of coupling in Figures 2a and 2b.

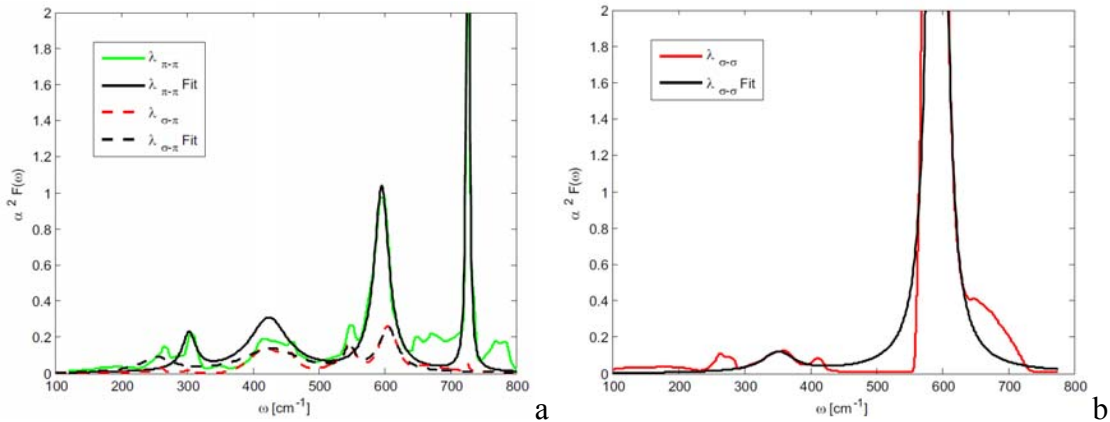

**Figure 2** Decomposition of Eliashberg function on different interaction types and construction of model function. Band contributions and corresponding fit to the model Lorenzian components; (a) $\lambda_{\pi\pi}$ and $\lambda_{\sigma\pi}$, (b) $\lambda_{\sigma\sigma}$. For details see main text. Conversion unit: 1 meV = 8.065441 cm$^{-1}$.

The $\lambda_{\pi\pi}$ function (Figure 2a) is fitted to four-component Lorenzian with the norm 0.448 (see Equation (D5) in Appendix D). The poles are at $\Omega_k$ = {303.1, 424.3, 595.7, 726.8} cm$^{-1}$ with appropriate heights and widths. Using the same procedure, the second important contribution $\lambda_{\sigma\pi}$ has been fitted with the norm 0.213 and poles at $\Omega_k$ = {254.8, 428.5, 546.8, 605.8} cm$^{-1}$. The crucial contribution, $\lambda_{\sigma\sigma}$, seems to have a large divergent peak near 600 cm$^{-1}$ (Figure 2b). This fact leads us to use a simpler 2-component model with poles at $\Omega_k$ = {351.6, 591.9} cm$^{-1}$, normalized to the value 1.017. The Eliashberg model functions obtained in this way have been used to construct the line-shape function that is crucial for the study of relaxation dynamics.

As discussed earlier, the source of electronic structure instability on the adiabatic level, which induces transition into a stabilized anti-adiabatic state, is electronic coupling to *ab*-plane B-B stretching vibration in E$_{2g}$ phonon mode with dominant frequency near to 600 cm$^{-1}$ (~74 meV). The *ab*-plane stretching is related to the $\lambda_{\sigma\sigma}$ and $\lambda_{\sigma\pi}$ type of coupling. The $\lambda_{\sigma\sigma}$ type, with the Eliashberg model function in Figure 2b, yields an opening of $\Delta_\sigma(0)$ gap (Figure 1c). A smaller gap, $\Delta_\pi(0)$, is opened in the $\pi$-band (Figure 1c), due to the $\lambda_{\sigma\pi}$ type of coupling, represented by the Eliashberg model function in Figure 2a. The couplings of $\lambda_{\pi\pi}$ and $\lambda_{\pi\sigma}$ type pertain to other phonon branches, that do not participate in the transition to an anti-adiabatic state, but are still important for relaxation dynamics over the whole temperature range. Until the system remains in a stabilized anti-adiabatic state, according to anti-adiabatic theory, the EP interaction energy given by Equation (8) for the phonon mode that is the source of electronic structure instability on the adiabatic level equals zero – see Equation (9a). This concerns the coupling to E$_{2g}$ phonon mode below T$_c$.

In the numerical modeling of relaxation dynamics, this means that, effective coupling $\lambda_{\sigma\sigma}$ = 0 for 0 < T ≤ T$_{c\sigma}$, and $\lambda_{\sigma\pi}$ = 0 for 0 < T ≤ T$_{c\pi}$. For the intermediate temperature range T$_{c\pi}$ < T ≤ T$_{c\sigma}$, within relaxation dynamics participates the coupling of $\lambda_{\sigma\pi}$, $\lambda_{\pi\pi}$ and $\lambda_{\pi\sigma}$ type. In the range 0 < T ≤ T$_{c\pi}$, only $\lambda_{\pi\pi}$ and $\lambda_{\pi\sigma}$ contribute to relaxation dynamics through the corresponding type of EP coupling: $\alpha^2 F_{\pi\pi}(\omega)$, $\alpha^2 F_{\pi\sigma}(\omega)$. For temperatures T > T$_{c\sigma}$ all types of EP coupling, i.e. $\alpha^2 F_{\sigma\sigma}(\omega)$, $\alpha^2 F_{\sigma\pi}(\omega)$, $\alpha^2 F_{\pi\pi}(\omega)$, $\alpha^2 F_{\pi\sigma}(\omega)$, contribute to relaxation dynamics.

In contrast to anti-adiabatic theory, BCS-like theories of superconductivity are formulated on the strict assumption of the validity of the adiabatic condition, as well as in a superconducting state. In these circumstances, Cooper-pair formation related to "BCS gap" opening is crucial, $\Delta=2\Delta_{BCS}$, at corresponding T$_c$. The temperature dependence of the gap formally obeys the same gap-equation, as it holds for an anti-adiabatic state transition, viz. Equations (6) and (7); only instead of the numerical factor of 4, the BCS-value of 3.53 is used. For modeling of relaxation dynamics, an important fact is that within the BCS model, all coupling types of EP interaction participate over the whole temperature range, no matter if the gap is opened or not. Switching off some EP coupling type below T$_c$ might prevent Cooper pair formation and destroy the superconducting state, in this case. Differences in the EP coupling scheme for numerical modeling of relaxation dynamics within AAT and BCS model are specified in Table 1.

| T-range | AAT model Participation of coupling type | | | | BCS model Participation of coupling type | | | |
|---|---|---|---|---|---|---|---|---|
| | $\lambda_{\sigma\sigma}$ | $\lambda_{\sigma\lambda}$ | $\lambda_{\pi\pi}$ | $\lambda_{\pi\sigma}$ | $\lambda_{\sigma\sigma}$ | $\lambda_{\sigma\lambda}$ | $\lambda_{\pi\pi}$ | $\lambda_{\pi\sigma}$ |
| $0<T\leq T_{c\pi}$ | N | N | Y | Y | Y | Y | Y | Y |
| $T_{c\pi}<T\leq T_{c\sigma}$ | N | Y | Y | Y | Y | Y | Y | Y |
| $T>T_{c\sigma}$ | Y | Y | Y | Y | Y | Y | Y | Y |

**Table 1** Electron-phonon coupling scheme: participation (Y/N = Yes/No) of particular interaction types in modeling relaxation dynamics for different temperature ranges and models.

### 3.3. Derived form of model PP signal

As specified in Appendix A, the PP experiment can be described as a third-order nonlinear process, when the resulting signal is directed by the probe wave-vector. The pump pulse itself causes two interactions, i.e. there are two time-contracted pump pulses with the same wave-vectors but in opposite directions, $\tau_1 = \tau_2; k_1 = -k_2$. The wave vector of probe pulse is in direction $k_3$. The detector measures the time-integrated intensity for times after the probe pulse acts. For a complex PP signal [29,30] with delay time $t_D$ between pump and probe can be written:

$$S(t_D) = \int_0^\infty P^{(3)}(t, t_D)dt = \int_0^\infty S_C(t, t_D)dt \cdot \quad (11)$$

Integration in Equation (11) is performed for $t$ times after the action of the probe pulse. In what follows, we use for $P^{(3)}(t, t_D)$ the term *complex signal $S_C(t,t_D)$*, i.e. $S_C(t, t_D) \equiv P^{(3)}(t, t_D)$. According to the relative phase of the pump and probe pulses,[29] the real signal is either the real part of $S(t_D)$, as we have used, or can be proportional to imaginary part of $S(t_D)$ – i.e. phases are shifted by an angle $\pi/2$.

The results presented below are for an OPOP setting of PP experiments with the assumption of a delta-pulse form of the temporal envelope of interacting fields, $E_i(t) = |E_i|\delta(t - t_0)$. The rotating-wave approximation (RWA) has been applied for elimination of fast oscillating optical terms $\exp(\pm i 2\omega_i t)$ in the evaluation of final form of $P^{(3)}$ - Equations (A11) and (C5) in the Appendices A and C. For description of EP interactions, we have applied a multimode quantum Brown oscillator model that allows the corresponding FWM correlation functions, $\{R_\alpha\}$ terms, see Equations (A14a-d), to be written in time-multiplicative form (Equations (11) and (A28)) of electronic and vibration contributions, which we have extracted from corresponding Eliashberg functions.

Within these circumstances, omitting some unimportant pre-factors, we can write the complex signal $S_C(t,t_D)$ for MgB$_2$ relaxation dynamics in the form,

$$S_C(t, t_D) = \begin{cases} R_2(t, t_D) \exp[-g^*(t) + 2i\,\mathrm{Im}(g(t_D) - g(t + t_D)) - t_D/T_1] \\ + R_3(t, t_D) \exp[-g(t)] \end{cases} \chi(t) \exp(i\Delta t) \cdot \quad (12)$$

Here, the $R_2(t, t_D), R_3(t, t_D)$ terms are the corresponding electronic parts of the response, i.e. *electronic dipole response functions* of $R_i(t_3, t_2, t_1)$: see Equations (A14b-c), with explicit form given in Equation (C5), derived for the PP experimental set-up:

$t_3 \to t$, $t_2 \to t_D$, $t_1 \to 0$, $\tau_1 = \tau_2$, $k_1 = -k_2$. In this case, out of four possible $\{R_i(t_3,t_2,t_1)\}$ terms given in Equations (A14a-d), only the $R_2$ and $R_3$ terms are independent, which are relevant for evaluation of $P^{(3)}(t)$. The microscopic processes that the $R_2$ and $R_3$ terms describe are visualized by time-evolution diagrams of perturbation theory in Appendices B1 and B2. For interpretation of the experimental results, the processes running over delay time period $t_D \equiv t_2$ are crucial. From Appendix B1, one can see that experimentally detected dynamics concern the time evolution of excited state population $\rho_{ee}(t_D)$, and represent the process of stimulated emission (SE) of the excited state as covered by $R_2$ term contributions. Within the simplified electronic model of MgB$_2$ (Figure 1d), the excited state $|e\rangle$ is represented by a single energy level and, consequently, the excited state absorption terms are not present. From Appendix B2, processes running over the delay time period $t_D \equiv t_2$, which are described by $R_3$ terms, are different, and represent processes of ground-state bleaching (GSB), either through depletion of the ground state population $\rho_{g_ig_i}(t_D)$ or through relaxation of ground-state coherences $\rho_{g_ig_j}(t_D)$.

In Equation (12), $\Delta = \omega - \omega_{eg}$ means detuning of the PP pulse central frequencies from the electronic excitation gap. The line-shape function $g(t)$ ($g^*(t)$ is the complex conjugate) describes the influence of dissipative environment, i.e. phonon bath. In order to include different phonon branches and different types of EP interactions for construction of the line shape function, we have used a multimode quantum Brownian oscillator model with Eliashberg functions [19] and its model simplification, as specified in section 3, with Meier-Tannor Lorenzian decomposition [26,27]. All necessary details on electronic correlation functions for studied model are in Appendix C. The form of the vibration part of correlation functions and line shape function, its derivation from Eliashberg functions, Meier-Tannor decomposition of spectral density function for multimode quantum Brownian oscillator model and Padé spectrum decomposition of the Bose function, $1/(1-e^{-\beta\omega})$, are all specified in Appendix D. Detailed analysis of the vibration correlation function has been performed for a single-mode/single-component Tannor model, which is presented in Appendix E.

The term $\exp(-t_D/T_1)$ in Equation (12) represents relaxation dynamics of the SE process, while $T_1$ is the excited state net fluorescence life-time, and corresponds to the longitudinal decay time, well known from Bloch model dynamics [29] – see Appendix C. The last term in Equation (12), $\chi(t)$, mimics the ensemble average over the statistical distribution of $\omega_{eg}$ (i.e. Inhomogeneous Line Broadening [29], IHLB), and we have used its standard Gaussian form, $\chi(t) = \exp(-t^2/2\sigma^2)$.

Finally, it should be remarked that the measured signal is a time-convolution of the net PP signal, $S(t_D)$, and an instrument response function G(t),

$$S_{ex}(t_D) = \int_0^\infty S(t_D - \tau)\, G(\tau)\, d\tau \; . \tag{13}$$

In our simulations we have used a Gaussian form of G(t).

## 4. Results of numerical modeling of $MgB_2$ relaxation dynamics
### 4.1. Line-shape function calculation

For the calculation of the PP signal, the spectral density $J(\omega)$, which enters the expression for the line-shape function calculation (Equations (D4a) and (D4b) in Appendix D) needs to be constructed. As a rule, we use the most simple linear relation, $J(\omega) = K \, \omega \, \alpha^2 F(\omega)$, i.e. spectral density is proportional to the Eliashberg function ($K$ is an appropriate numerical pre-factor). The spectral density has been decomposed into Meier-Tannor form [26,27] of Lorentzians (Appendix D) and, the previously derived parameters of Lorenzian decomposition of $\alpha^2 F(\omega)$ are now used as the input in the line-shape function calculation.

The temporal behavior of the calculated line-shape function is presented in Figure 3.

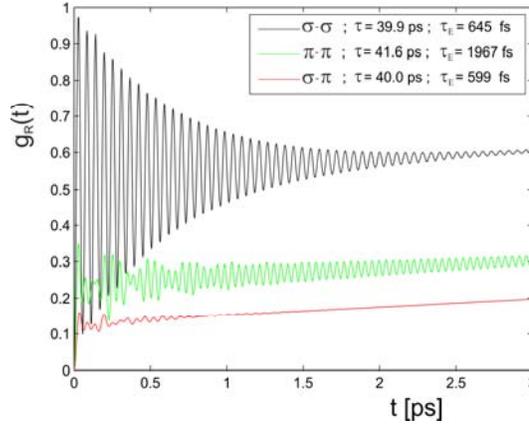

**Figure 3** Temporal evolution of the real part of the line-shape functions for a particular interaction type at temperature 5 K.

It should be remarked that, for spectral density in the PHz frequency range, we have used the auxiliary pre-factor $K$ in order to obtain the resulting line-shape function decay times in the few-ps time region (Figure 4b). For each coupling type, the temperature-dependent line-shape function was constructed, but with respect to differences in coupling specification for the studied AAT and BCS models, as it has been introduced in section 3.2 (Table 1). We note that the critical temperatures for both studied models have been assumed to be the same, viz. $T_{c\pi} \approx 16.2$ K and $T_{c\sigma} \approx 40.6$ K – according to Equation (7).

In the calculation of the integrated signal, Equation (11), it is desirable to know the overall behavior of the line-shape function, especially its real part $g_R(t)$. By inspection of Figure 3, we can anticipate that for each coupling type (as well for multimode coupling), the temporal behavior of $g_R(t)$ can be described in a general form (for details see Appendix D and E),

$$g_R(t) = a + t/\tau + e^{-t/\tau_E} osc(t), \qquad (14)$$

where $\tau$ has the meaning of long-time decay and $\tau_E$ determines the lifetime of oscillating part. The term $osc(t)$ in Equation (14) stands for the oscillating part, which for the studied system consists of contributions with periods ~ 50–120 fs. The symbol $a$ is a constant. Knowledge of $\tau$ is desirable for the numerical integral calculation, see

Equations (11) and (12), since both terms of the response function, Equation (12), contain the pre-factor $\exp(-t/\tau)$.

The results of the temperature dependence of long time decay, $\tau$, of the multimode $g_R(t)$ for two studied models are displayed in Figure 4a. Temperature dependence of decay $\tau_E$ of oscillating part of the line-shape function is shown in Figure 4b. From Figure 4a one can see that in the low temperature regime (below $T_{c\pi}$) the two models possess diametrically different decay times, which is due to the presence of intense $\sigma-\sigma$ coupling contribution (see Figure 3) in the BCS model. For temperatures above $T_{c\pi}$, the difference in decay times is not substantial, and above $T_{c\sigma}$, the decay times for both models are identical. For oscillating decay (Figure 4b) up to the temperature $T_{c\sigma}$ a dramatic difference exists between the two models. As far as we know, this form of $\tau_E$ is nearly independent of the auxiliary line-shape constant $K$ and based on our analysis in Appendix E, Equations (E13–E16), it depends on an "average" width $\Gamma$ of a spectral density peak.

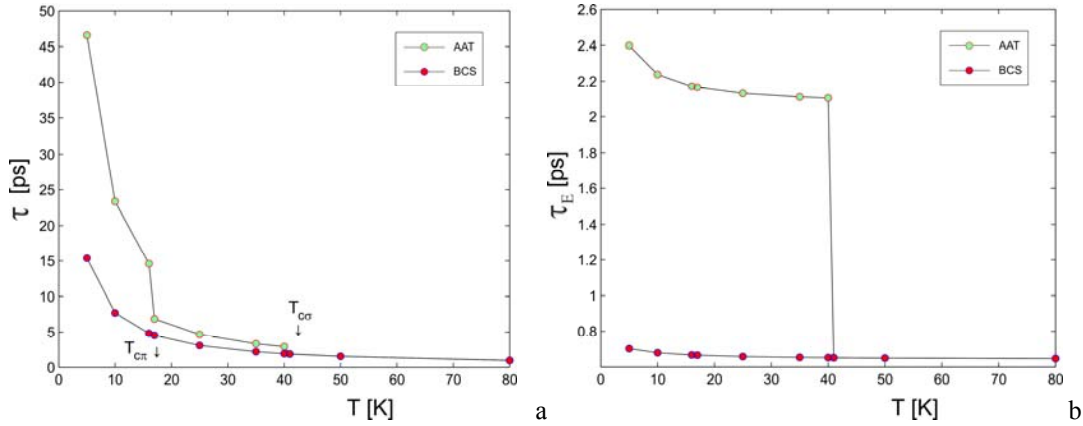

**Figure 4** Temperature dependence of (a) long-time decay and (b) oscillating decay of the real part of the multimode line-shape function g(t) for the studied AAT and BCS models. Critical temperatures are indicated by arrows.

### 4.2. Calculation of the PP signal

According to Equations (11) and (12) we have calculated the integrated PP signal and its convolved counterpart, Equation (13). The excited state net fluorescence life-time $T_1$, was set to *ps* time-range. In our numerical calculations, the value $T_1 = 10$ ps has been used. In order not to interfere with the IHLB influence, a temporal width of $\chi(t)$ in Equation (12) was set to 20 ps. The PP signal was calculated in the range 0–50 ps with proper selection of the upper value of the numerical integral (it was set to $3.5\,\tau$, see Figure 4b) and with an integration step of 2.5 fs. The detuning value was set to zero (Equation (12): $\Delta = 0$). Finally, the real part of $S(t_D)$ in Equation (11), i.e. $Re[S(t_D)]$, has been selected as a net signal.

As an illustration, numerically obtained signals as a function of delay time $t_D$ for the AAT model, calculated according to Equation (11) for different temperatures, are displayed in Figure 5a.

The net PP signal, calculated from Equation (11), contains information on the decay processes that are involved. To estimate the parameters of relaxation dynamics, each signal was fitted to the double-exponential form, $a_1 e^{-t_D/\tau_1} + a_2 e^{-t_D/\tau_2}$ ($t_D$ is pump-probe delay time).

Figure 5b illustrates the temperature dependence of the $\tau_1$ decay-time that we have obtained for the AAT and BCS models.

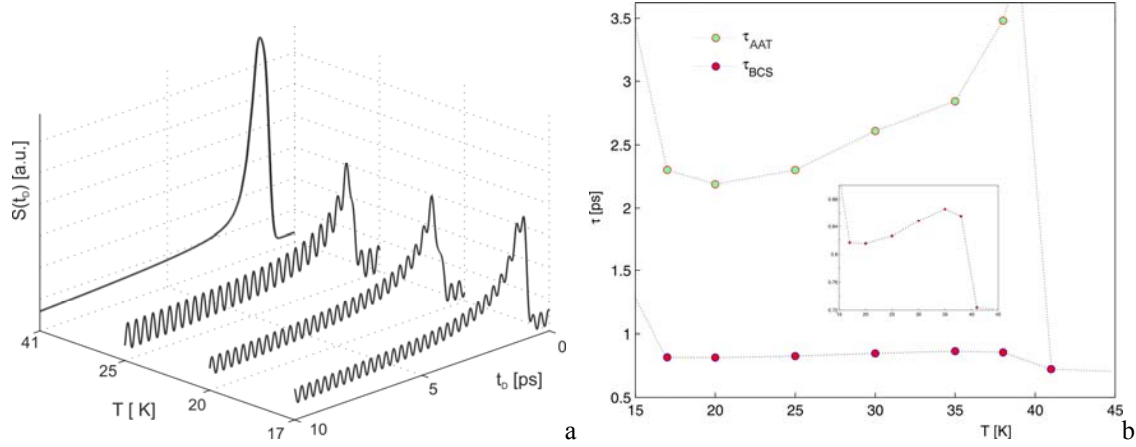

**Figure 5** (a) Time dependence of the convolved PP signal for an anti-adiabatic AAT model at different temperatures (17 K, 20 K, 25K, 41 K). The net signal was convolved with a 200 fs Gaussian instrument response according to Equation (13). (b) Temperature dependence of decay time of the net PP signal for studied AAT and BCS models. The inset shows a zoom of the decay times for BCS model.

As it can be seen from Figure 5b, above $T_{c\sigma}$ ($\approx$ 40 K), relaxation dynamics in both studied models is identical. This corresponds to relaxation dynamics of an effective two-level system.

For the AAT model in a cooling regime, as soon as $T_{c\sigma}$ is reached, the relaxation time suddenly increases with respect to relaxation dynamics above $T_{c\sigma}$ and at 39 K it is still ~ 21 times larger. With a further decrease of temperature below $T_{c\sigma}$, the relaxation time smoothly decreases, and close to $T_{c\pi}$, a sudden increase of relaxation time starts again. Below $T_{c\pi}$, an increase of relaxation time continues, according to "standard" temperature decrease of the population of phonon modes.

For the BCS model, the temperature dependent behavior of the relaxation time is much less striking than in the case of the AAT model. Nevertheless, in the enlarged scale of $\tau(T)$ for BCS model (inset in Figure 6), one can see that close to $T_{c\pi}$ and $T_{c\sigma}$ when gaps in the single-particle spectrum are opened, sudden, but relatively small, changes in relaxation time can also be identified.

## 5. Discussion and conclusions

The present study is the first attempt to model pump-probe relaxation dynamics of superconductors on the basis of a first-principles formulation of nonlinear optics. Numerical outputs are always dependent on the studied model and on the accuracy and suitability of the applied numerical methods. The results presented in this study for AAT model are in good agreement with experimental OPOP dynamics [10] as far as the temperature dependence of relaxation time on a few-ps time scale is concerned. Numerical modeling yields abrupt changes of relaxation time near to 40 K and 16 K. These are critical temperatures for opening the gaps in single particle-spectrum in the σ and π-band of MgB$_2$. Abrupt changes of relaxation time close to these temperatures are the consequence of sudden changes in the character of EP coupling in the transition from an adiabatic to a stabilized superconducting anti-adiabatic state, as it predicts the anti-adiabatic theory of electron-vibration interaction for MgB$_2$. The sole effect of opening the gaps and corresponding T-dependence of the gaps evolution is not strong enough for the observed enormous increase of relaxation time at 40 K and 16 K. The BCS model, which preserves the adiabatic character of EP coupling also below the critical temperature of MgB$_2$, is basically unable to reflect this enormous sudden increase of relaxation time.

From the stand point of the application of first-principles theory to PP spectroscopy of superconductors, certain aspects are important, which allow identification of non-equilibrium processes running on a microscopic level over different time periods before a coherent PP signal is detected, depending on the delay-time $t_D$ between pump and probe pulses. In the case of a OPOP experiment, from the derived form of the 3$^{rd}$-order response function and the diagrammatic representation (Appendix B) of Liouville path-evolution dynamics, it follows that the relaxation process detected as $t_D$-evolution of an emitted coherent polarization field is dominated by stimulated emission (SE – $R_2$ term) of excited state. It is a relaxation of the excited state population $\rho_{ee}$ expressed through the fluorescence life time, i.e. longitudinal relaxation time T$_1$. Processes of ground-state bleaching (GSB), i.e. mainly relaxation of coherences in the ground state, $\rho_{gigj}$, modulate to some extent the relaxation dynamics, through contributions of electronic correlation functions in $R_3$ term. Since for modeling, the δ-form of the temporal-envelope of incident laser pulses has been used (the non-adiabatic regime of field-matter interactions), the calculated time dependence of relaxation dynamics has the form of dumped oscillating relaxation dynamics, mainly at low temperatures. Oscillating dynamics in non-adiabatic regimes were also recently reported [14,16,17] for the time evolution of the BCS order parameter calculated depending on the duration of a pump pulse.

The other point of the present microscopic insight into PP relaxation dynamics of superconductors relates to aspects of EP interactions in relation to the decay time of relaxation. The dimensionless EP coupling constant $\lambda$ is directly related to the Eliashberg function through the second moment, $\lambda \langle \omega^2 \rangle = 2 \int_0^\infty d\omega (\alpha^2 F(\omega)) \omega$ and represents integral norm, $\lambda = 2 \int_0^\infty d\omega (\alpha^2 F(\omega)) / \omega$. Term $\lambda \langle \omega^2 \rangle$ is directly related [4] to relaxation time due to EP coupling, $\tau_{ep}$, through the relation $\lambda \langle \omega^2 \rangle = \frac{\pi}{3} \frac{k_B T_e}{\hbar \tau_{ep}}$, with $T_e$ representing temperature of "hot" electrons, i.e. temperature of electrons immediately after excitation. Through the non-equilibrium model [8], this relationship has been derived in the form $\lambda \langle \omega^2 \rangle = \frac{2\pi}{3} \frac{k_B T_l}{\hbar \tau_{ep}}$, where $T_l$ is lattice temperature, which can be controlled better than $T_e$. Moreover in this case, it is not necessary to assume that electron-electron scattering is much faster than EP scattering. The coupling constant $\lambda$ is a key parameter of all BCS-like theories of superconductivity, directly related to the critical temperature of superconductor, $k_B T_c = \hbar\omega_0 \exp[-(1 + \lambda)/\lambda]$ (here, the pseudopotential $\mu^*$ is neglected). This offers a possibility to extract the EP coupling constant $\lambda$ through the relaxation time $\tau_{ep}$ from experimental data [41] of excited state relaxation dynamics. In assessment of the role of EP interactions in the mechanism of superconducting state transition, the replacement $\lambda\langle\omega^2\rangle \to \lambda\langle\omega_0^2\rangle$ is used, while $\omega_0$ is some characteristic frequency that is assumed to be crucial for transition into superconducting state.

Analysis of vibration correlation functions, which we have presented in Appendix D and in more detail in Appendix E, indicates that extraction of EP coupling constant $\lambda$ from experimental data of excited state relaxation dynamics is a more complex problem. Even in the case when decomposition of the Eliashberg function on particular interaction types is known, identification of a characteristic frequency $\omega_0$ is not trivial. There are two issues here. Relaxation dynamics, even if represented by a single mode-single component Eliashberg function with a peak at $\omega_0$, does not depend only on the peak frequency $\omega_0$, but also on the frequency distribution, i.e. the width of the peak $\Gamma$ and the peak-weight $p$: see Equations (E14–E17). The replacement $\lambda\langle\omega^2\rangle \to \lambda\langle\omega_0^2\rangle$ means that $\lambda$ extracted from relaxation dynamics in this way is an extreme simplification, where the dimensionless coupling constant $\lambda \equiv \lambda_{HR}$, whilst $\lambda_{HR}$ is a simple model EP coupling known under the term Huang-Rhys factor: Equations (E21,E22). Moreover, a line-shape function constructed on the basis of δ-peak frequency spectral model, even if the discrete sum of the δ-peak frequencies is considered, cannot describe the temperature dependence of relaxation dynamics – see Equation (E21).

The other aspect, which complicates the situation, is the fact that a single phonon branch can induce several types of EP coupling (e.g. $E_{2g}$ in $MgB_2$ induces $\sigma\sigma$ and $\sigma\pi$-type) and each type of coupling is characterized by a multicomponent EP coupling structure, i.e. by several peaks at different frequencies. It can be easily shown that in case of $\sigma\sigma$-type of coupling, which drives transition into the superconducting state of $MgB_2$, in order to recover the EP coupling constant $\lambda_{HR\sigma\sigma} \approx \lambda_{\sigma\sigma} \approx 1$ as it corresponds to the Eliashberg function (Figure 2b), the "characteristic" frequency has to be $\omega_0 \approx 21$ cm$^{-1}$. This value corresponds to the frequency region of acoustic phonons, far away from the dominant frequency of optical $E_{2g}$ phonon mode that drives the transition into a superconducting state. On the other hand, if we assume that the characteristic frequency is either $\omega_0 \approx 592$ cm$^{-1}$ or $\omega_0 \approx 362.8$ cm$^{-1}$ as it corresponds to Eliashberg function (Figure 2b), then the calculated coupling constant is $\lambda_{HR} = 0.153$ or $\lambda_{HR} = 0.007$, respectively. The resulting $\lambda_{HR\sigma\sigma} = 0.16$ is too small to reproduce $T_c \approx 40$ K within a BCS-like theory. For high-temperature cuprates, the situation can be yet more complicated. In the case of YBCO, for instance, it can be expected that for the transition into a superconducting state, the participation of three phonon branches is crucial [31]. In particular, it concerns the $B_{2g}$, $B_{3g}$ modes of Cu-O$_2$(O$_3$) stretching vibrations and the vibration of apical O$_4$ in $A_g$ mode.

Nevertheless, a direct relationship with the mechanism of superconducting state transition can be traced from experimental PP relaxation dynamics. As it follows from the AAT model of the present study, change in decay time for temperatures near to $T_c$, where an abrupt change in relaxation dynamics is detected, e.g. $\Delta\tau = \tau(T_c - 1) - \tau(T_c + 1)$, is the direct consequence of the "decoupling" of EP interactions mediated by phonon mode(s) which is (or are) responsible for superconducting state transition. When the decomposition of the Eliashberg function to the contributions of participating phonon branches is available, their dynamics can be calculated. Phonon mode, or particular type of interaction within this phonon mode, which in "decoupling" yields a sudden change in relaxation time $\Delta\tau_{cal}$ comparable to the experimental change $\Delta\tau_{exp}$, should be the mode responsible for superconducting state transition. In the case of cuprates, it can be a combination of several modes, as has been mentioned before.

Finally, differences in OPOP vs. OPTP relaxation dynamics of $MgB_2$ along with aspects of relaxation time detected over a delay-time period $t_D$ and the recovery of superconducting state should be briefly discussed. As can be seen from the Liouville path time evolution diagrams in Appendices B3 and B4 for the OPTP setting, dynamics detected as a function of $t_D$ pertains to the time-evolution of different microscopic states as compared to the OPOP (Appendices B1 and B2) experiment. The stimulated emission of excited state $|e\rangle\langle e|$ (stimulated fluorescence) in OPTP is not detected over delay-time period $t_D$, in contrast to the OPOP setting where this process is dominant for relaxation dynamics – c.f. diagrams B1–B4. What is crucial for OPTP dynamics are the processes of relaxation of the ground state coherences $|g_i\rangle\langle g_j|$. Relaxation processes within the ground state (involving energy levels {1, 2, 3}) are driven by phonon-phonon interactions and the quantum Brownian oscillator model, which has been applied for OPOP, is not possible to apply so straightforwardly for system-bath interactions in this case. It is expected that the relaxation process will be controlled now by the interaction of the anharmonic $E_{2g}$ mode with the rest of the phonon branches in $MgB_2$. This approach will be presented elsewhere. Nevertheless, one important aspect of OPTP dynamics can be extracted from the Liouville path time evolution diagrams. Last two diagrams in B4 (Figure B4b) demonstrate a possibility that spontaneous relaxation of the excited state $|e\rangle\langle e|$ (spontaneous florescence) contributes to relaxation dynamics detected over delay-time period $t_D$ in the OPTP setting. In general, spontaneous fluorescence is a slower process than stimulated fluorescence, and the relaxation process detected by OPTP setting can be expected to be slower (e.g. as reported in [11]) than the relaxation detected within OPOP experiments.

It should be remarked that the relaxation time $\tau$ extracted from the dynamics over a delay-time period $t_D$ does not mean that, immediately after the detection of an emitted field, a superconducting state in its thermal equilibrium with the surrounding environment is recovered. From all the Figures in Appendix B, one can see that the final state after emission of polarization field leaves the system in the ground state, but different energy levels of the ground state are populated, $\{|1\rangle\langle 1|, |2\rangle\langle 2|, |3\rangle\langle 3|\}$. It is obvious that the populations $\{|2\rangle\langle 2|, |3\rangle\langle 3|\}$ have to relax for some future time period until the superconducting state $|1\rangle\langle 1|$ in its thermal equilibrium is recovered. In this respect, an interesting possibility is indicated by Figures B3 and B4 for temperature range $T_{c\pi} < T \leq T_{c\sigma}$ in the OPTP setting. In these cases, after emission of a polarization field, the system is in $|1\rangle\langle 1|$ final state, and the relaxation time $\tau$ extracted from the dynamics over delay-time period $t_D$ should represent the relaxation time of recovery of a superconducting state that is in thermal equilibrium with the environment.

In conclusion, the theoretical method based on a first-principles formulation of nonlinear optics represents an effective tool in modeling pump-probe relaxation dynamics of superconductors. Knowledge of the electronic structure of the studied system and the corresponding Eliashberg function that represents pertinent EP interactions enables us to distinguish nonequilibrium processes running over different time-periods in a sequence of interactions with laser pulses on a microscopic level. We have also derived the temperature dependent relaxation dynamics as a function of delay time between pump and probe pulses. For the studied model system of $MgB_2$, it has been shown that an abrupt increase of relaxation time at $T_c$, as detected experimentally, is the direct consequence of sudden changes in the character of EP coupling in transition from an adiabatic to a stabilized superconducting anti-adiabatic state, as predicted from anti-adiabatic theory of electron-vibration interactions. The BCS model, which preserves the adiabatic character of EP coupling also below the critical temperature of $MgB_2$, is basically unable to reflect this enormous sudden increase of relaxation time.

**Acknowledgements**
This work was supported by the Slovak Research and Development Agency under the contract No. APVV-0201-11.

**Appendix A**
**Density matrix formulation of 3$^{rd}$-order polarization in modeling of PP spectroscopy**
Optically induced polarization is a source of a new field irradiated by a sample as a signal $E_{sig} \propto P$. From Maxwell equation for longitudinal component of field interacting with a matter,

$$\nabla^2 E(r,t) - \frac{1}{c^2}\frac{\partial^2 E(r,t)}{\partial^2 t^2} = \frac{4\pi}{c^2}\frac{\partial^2 P(r,t)}{\partial^2 t^2}, \tag{A1}$$

solution for polarization can be derived in the form,

$$P(r,t) = P(t)\exp(ik_{sig}\cdot r - i\omega_{sig}t) + c.c. \tag{A2}$$

In order to satisfy energy and momentum conservation, the wave vector and frequency of polarization have to obey the relations,

$$k_{sig} = \sum_i \pm k_i, \quad \omega_{sig} = \sum_i \pm \omega_i \tag{A3}$$

with respect to wave vectors $\{k_i\}$ and frequencies $\{\omega_i\}$ of incident external electric fields $\{E_i(r,t)\}$,

$$E_i(r,t) = E_i \exp(-i\omega_i t + ik_i r) + c.c. \tag{A4}$$

In these circumstances, oscillating polarization radiates coherent signal $E_{sig}$ with frequency $\omega_{sig}$ in wave vector direction $k_{sig}$.

In theoretical modeling, polarization can be expressed as the mean value of dipole moment $\langle \mu(t) \rangle$, i.e.

$$P(t) = Tr(\mu\rho(t)), \tag{A5}$$

and can be calculated through density matrix, which obeys the Liouville equation,

$$\frac{d\rho(t)}{dt} = -\frac{i}{\hbar}L\rho(t) = -\frac{i}{\hbar}[H,\rho(t)] \tag{A6}$$

with Liouvillian $L$ that pertains to Hamiltonian $H$, of studied system.

In optical spectroscopy, Hamiltonian H of studied system is composed of unperturbed part $H_0$ and of perturbation $H_{int}$, due to interaction with an external incoming light field,

$$H = H_0 + H_{int}(t) \tag{A7}$$

In dipole moment approximation,

$$H_{int}(t) \equiv V(t) = -\mu E(t) \tag{A8}$$

The field only serves to induce transitions between quantum states of studied system. With respect to internal Coulomb forces acting inside the system, interaction with external field can be treated as a small perturbation and $n^{th}$-order perturbation theory can be used if system interacts with a sequence of $n$ external fields, which is the case of nonlinear optical spectroscopy. In these circumstances, polarization can be expanded in powers of interacting fields,

$$P(t) = P^{(0)} + P^{(1)} + P^{(2)} + P^{(3)} + \ldots = Tr(\mu\rho^{(0)}(t)) + Tr(\mu\rho^{(1)}(t)) + Tr(\mu\rho^{(2)}(t)) + Tr(\mu\rho^{(3)}(t)) + \ldots \tag{A9}$$

Relevant for FWM spectroscopy is $3^{rd}$-order polarization; $P^{(3)}(t) = Tr(\mu\rho^{(3)}(t))$. Expansion of density matrix (A6) in $3^{rd}$-order of interacting fields results in,

$$\rho_I^{(3)} = \left(-\frac{i}{\hbar}\right)^3 \int_{-\infty}^{t} d\tau_3 \int_{-\infty}^{\tau_3} d\tau_2 \int_{-\infty}^{\tau_2} d\tau_1 [V_I(\tau_3),[V_I(\tau_2),[V_I(\tau_1),\rho_{eq}]]] \tag{A10}$$

Subscript $I$ in equation (A10) stand for interaction picture representation and $\tau_i$ $\{i = 1,2,3\}$ is an instant of time when particular field $E_i$ $\{i = 1,2,3\}$ hits a system. In the far-past $(-\infty)$, till interaction with the first incoming field, system is in thermal equilibrium, $\rho_{eq} = \exp(-\beta H_0)/Tr(\exp(-\beta H_0))$.

For $P^{(3)}(t)$ then follows [29],

$$P^{(3)}(t) = \int_0^\infty dt_3 \int_0^\infty dt_2 \int_0^\infty dt_1 S^{(3)}(t_1,t_2,t_3) E_1(t-t_1-t_2-t_3) E_2(t-t_2-t_3) E_3(t-t_3) \tag{A11}$$

with $3^{rd}$-order response function $S^{(3)}$,

$$S^{(3)}(t_1,t_2,t_3) = \left(\frac{i}{\hbar}\right)^3 \theta(t_1)\theta(t_2)\theta(t_3) Tr\{[[[\mu_I(t_3+t_2+t_1),\mu_I(t_2+t_1)],\mu_I(t_1)],\mu_I(0)]\rho_{eq}\} \tag{A12}$$

Form of equation (A11) is expressed now through consecutive time intervals $\{t_1 \to t_2 \to t_3\}$ prior to detection - observation of irradiated field. It is based on substitutions; $t_1 = \tau_2 - \tau_1$, $t_2 = \tau_3 - \tau_2$, $t_3 = t - \tau_3$, whilst $\theta(t_{i=1,2,3})$ is Heaviside step function. First pulse interacts with sample at instant of time $\tau_1 = t_0 = 0$ and, $\rho(t_0) = \rho(0) = \rho_{eq}$. Evaluation of nested commutators in equation (A12) yields four distinct sets of multi-point correlation functions of dipole moment and $S^{(3)}$ can be written in the form,

$$S^{(3)}(t_1,t_2,t_3) = \left(\frac{i}{\hbar}\right)^3 \theta(t_1)\theta(t_2)\theta(t_3) \sum_{\alpha=1}^{4}(R_\alpha - R_\alpha^*) \tag{A13}$$

where $R_\alpha^*$ is complex conjugate of $R_\alpha$.

Particular $\{R_\alpha\}$ terms are,

$$R_1(t_1,t_2,t_3) = Tr(\mu(t_1)\mu(t_1+t_2)\mu(t_1+t_2+t_3)\mu(0)\rho(0)) \tag{A14a}$$

$$R_2(t_1, t_2, t_3) = Tr(\mu(0)\mu(t_1 + t_2)\mu(t_1 + t_2 + t_3)\mu(t_1)\rho(0)) \tag{A14b}$$

$$R_3(t_1, t_2, t_3) = Tr(\mu(0)\mu(t_1)\mu(t_1 + t_2 + t_3)\mu(t_1 + t_2)\rho(0)) \tag{A14c}$$

$$R_4(t_1, t_2, t_3) = Tr(\mu(t_1 + t_2 + t_3)\mu(t_1 + t_2)\mu(t_1)\mu(0)\rho(0)) \tag{A14d}$$

Time-dependence of dipole moment operator, $\mu(t) = U_0^+(t)\mu U_0(t)$, is evaluated through time evolution operator $U_0 = \exp(-\frac{i}{\hbar}H_0 t)$ with respect to unperturbed Hamiltonian. It is obvious that interaction with external fields $\{E_i(r,t)\}$, mainly in a system with a multiple quantum states, can results in great number of permutations and different $R$-terms can yield a wide spectrum of different nonlinear signals related to different microscopic states which evolve in a system and, signals (emitted radiations) may differ by frequencies and wave vectors.

To keep time-ordered sequence, $\tau_1 \leq \tau_2 < \tau_3$, the $R_\alpha$ $\{\alpha = 1,2,3,4\}$ terms strictly distinguish the *ket* or *bra* side of density matrix, $\rho(t) = \sum_{a,b} \rho_{ab}(t)|a\rangle\langle b|$, in interaction with external fields $\{E_i(r,t)\}$. For particular terms, the side-specified sequence of interactions is; $R_1 \Rightarrow$ *ket/ket/ket*, $R_2 \Rightarrow$ *bra/ket/bra*, $R_3 \Rightarrow$ *bra/bra/ket*, $R_4 \Rightarrow$ *ket/bra/bra*. Corresponding time-evolution diagrams of diagrammatic perturbation theory in Liouville space are then an effective way for fast-tracking of possible non-equilibrium states evolving over the sequence of interactions with external fields and theirs contributions into $P^{(3)}$ in a studied system. In the Appendix B, relevant diagrams for OPOP and OPTP setting of PP spectroscopy for a model system of MgB$_2$ are displayed.

For evaluation of multi-point correlation functions in (A14), it is advantageous to decompose unperturbed system Hamiltonian $H_0$ into relevant part, in our case it is electron subsystem $H_e$, and the rest – set of vibration modes of a system forming thermal bath, which is represented by Hamiltonian $H_B$,

$$H_0 = H_e + H_B + H_{eB} \tag{A15}$$

Mutual interaction of electronic subsystem and a bath represents Hamiltonian $H_{eB}$.

For practical applications, the gap Hamiltonian form of $H_0$ is introduced. It then yields the multimode quantum-Brownian oscillator model[29] of electron subsystem-bath interactions. In general,

$$H_0 = H_e + H_B + H_{eB} = H_g(P,Q)|g\rangle\langle g| + H_{ex}(P,Q)|e\rangle\langle e| \tag{A16}$$

Electronic subsystem is now written over eigenstate(s) of a ground state $|g\rangle$ and electronically excited state $|e\rangle$. Nuclear part of system Hamiltonian pertaining to electronic ground and excited state is expressed through nuclear momenta $P$ and coordinates $Q$ as a sum of nuclear kinetic $T(P)$ and potential energy $V(Q)$. Then,

$$H_0 = \{\varepsilon_q|g\rangle\langle g| + \{\varepsilon_e + \langle V_{ex}(Q) - V_g(Q)\rangle\}|e\rangle\langle e|\} + \{T(P) + V_g(Q)\} +$$
$$+ \{V_{ex}(Q) - V_g(Q) - \langle V_{ex}(Q) - V_g(Q)\rangle\}|e\rangle\langle e| = H_e + H_B + H_{eB} \tag{A17}$$

This form assumes validity of Condon approximation, i.e. independence of transition dipole moment on nuclear coordinates, $\mu = d_{eg}|e\rangle\langle g| + d_{eg}^*|g\rangle\langle e|$. Then, Hamiltonian of electron subsystem-bath interactions can be transformed into gap Hamiltonian form. This is expressed as a fluctuation of electronic excitation gap, i.e. fluctuation in difference of electronic eigenenergies of excited and ground state, $\hbar\omega_{eq} = \varepsilon_e - \varepsilon_q$, which is a function of nuclear coordinate $Q$, and represents measure of coupling strength of electronic subsystem to nuclear motion,

$$H_{eB} = \left\{ V_{ex}(Q) - V_g(Q) - \langle V_{ex}(Q) - V_g(Q) \rangle \right\} |e\rangle\langle e| = \Delta V(Q)|e\rangle\langle e| \tag{A18}$$

Since $H_0$ is diagonal, then for evolution operator holds,

$$U_0(t) = U_{g0}(t)|g\rangle\langle g| + U_{e0}(t)|e\rangle\langle e|, \tag{A19}$$

where,

$$U_{0g}(t) = \exp\left\{-\frac{i}{\hbar}(\varepsilon_g + H_B(P,Q))t\right\} = \exp(-\frac{i}{\hbar}\varepsilon_g t)\widetilde{U}_g(t); \tag{A20}$$

$$\widetilde{U}_g(t) = U_B(t) = \exp\left\{-\frac{i}{\hbar}H_B(P,Q)t\right\}$$

and,

$$U_{0e}(t) = \exp\left\{-\frac{i}{\hbar}(\varepsilon_e + H_B(P,Q) + \Delta V(Q))t\right\} = \exp(-\frac{i}{\hbar}\varepsilon_e t)\widetilde{U}_e(t) \tag{A21}$$

with,

$$\widetilde{U}_e(t) = \exp\left\{-\frac{i}{\hbar}(H_B(P,Q) + \Delta V(Q))t\right\} = \exp\left\{-\frac{i}{\hbar}(H_B(P,Q))t\right\}\widetilde{U}_{eI}(t)$$

In interaction picture, for evolution operator can be derived,

$$\widetilde{U}_{eI}(t) = \exp\left\{-\frac{i}{\hbar}\int_0^t U_B^+(\tau)\Delta V(Q)U_B(\tau)d\tau\right\} \tag{A22}$$

For product of vibration part of evolution operators then holds,

$$\widetilde{U}_g(t)\widetilde{U}_e(t) = \exp\left\{-\frac{i}{\hbar}\int_0^t U_B^+(\tau)\Delta V(Q)U_B(\tau)d\tau\right\} \tag{A23}$$

This expression enables to calculate dipole moment correlation function of studied system in a factorized form, as a product of electron and nuclear-vibration contributions, i.e.

$$\langle\mu(t)\mu(0)\rangle = |d_{eg}|^2 \exp(-\frac{i}{\hbar}\omega_{eg}t)Tr_Q\left\{\exp\left\{-\frac{i}{\hbar}\int_0^t U_B^+(\tau)\Delta V(Q)U_B(\tau)d\tau\right\}\rho_{eq}\right\} \tag{A24}$$

Cumulant expansion of exponential integral part of (A24) in the first-order equals zero and up to second-order can be derived,

$$Tr_Q\left\{\exp\left\{-\frac{i}{\hbar}\int_0^t U_B^+(\tau)\Delta V(Q)U_B(\tau)d\tau\right\}\rho_{eq}\right\} = 1 - 0 - \frac{1}{\hbar^2}\int_0^t d\tau \int_0^\tau d\tau' Tr_Q\left\{U_B^+(\tau')\Delta V U_B(\tau')\Delta V \rho_{eq}\right\} \tag{A25}$$

Introduction of "vibration correlation function",

$$C(t) = \frac{1}{\hbar^2}Tr_Q\left\{U_B^+(t)\Delta V(Q)U_B(t)\Delta V(Q)\rho_{eq}\right\} \tag{A26}$$

and *line-shape function* in the form,

$$g(t) = \int_0^t d\tau \int_0^\tau d\tau' C(\tau') \tag{A27}$$

allows to write dipole moment correlation function (A24) in a simple form,

$$\langle \mu(t)\mu(0) \rangle = |d_{eg}|^2 \exp(-\frac{i}{\hbar}\omega_{eg}t)\exp(-g(t)) \tag{A28}$$

Explicit form of $\{R_\alpha\}$-terms expressed through $g(t)$ can be found in [29,30] and terms relevant for model of present study are specified in Appendix D.

In fact, $C(t)$ is correlation function which characterizes electron-vibration interaction expressed through electronic excitation gap $\Delta V(Q)$, which is a function of generalized coordinate $Q$ of particular vibration. It can be shown straightforwardly that excitation gap can be written in the form $\Delta V(Q) = kQ_e Q = \gamma Q$, and for $C(t)$ then holds,

$$C(t) = \frac{\gamma^2}{\hbar^2} Tr_Q\{U_B^+(t)QU_B(t)Q\rho_{eq}\} = \frac{\gamma^2}{\hbar^2}\langle Q(t)Q(0)\rho_{eq}\rangle.$$

It is obvious that electron subsystem interacts not just with a single mode, but with a set of vibration (phonon) modes and $\{C(t), g(t)\}$ are then expressed as a discrete summation over independent modes. In case of continuous distribution of modes it is expressed as an integral through corresponding density of states $F(\omega)$. In that case, correlation function can be directly related to the spectral density $J(\omega)$ derived from particular Eliashberg function $\alpha^2 F(\omega)$ - see Appendix D.

## Appendix B
## Liouville path time evolution diagrams of $3^{rd}$-order response with delay time $t_D \equiv t_2$ for relaxation dynamics of $MgB_2$

**Appendix B1** The $R_2$ term contributions to $3^{rd}$-order response function - diagrammatic representation for OPOP setting of incident fields.

The PP experimental set-up: $t_3 \to t$, $t_2 \to t_D$, $t_1 \to 0$, $\tau_1 = \tau_2$, $k_1 = -k_2$, $k_3$. Pump pulses $E_1$, $E_2$ and probe pulse $E_3$ are optical; $|E_1| = |E_2| = |E_3|$, $\omega_1 = \omega_2 = \omega_3 \equiv \omega_P$. Time ordering of incident fields in $\boldsymbol{R_2}$ term with respect to density matrix is: *bra / ket / bra*

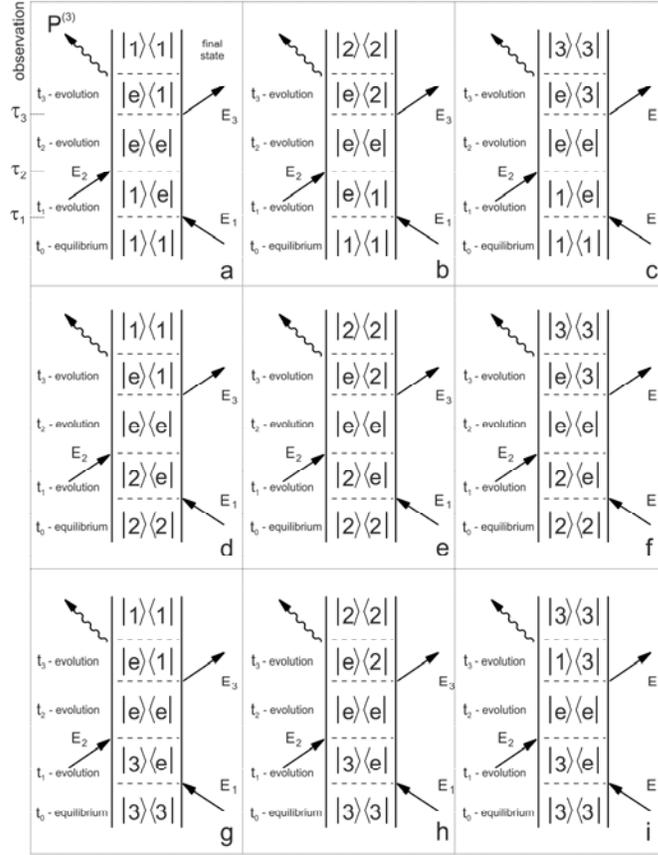

**Figure B1** The $R_2$ term contributions to response function for OPOP spectroscopy of $MgB_2$

In temperature range: $0 < T \leq T_{c\pi}$ - all diagrams contribute; $T_{c\pi} < T \leq T_{c\sigma}$ - diagrams (a,c,g,i) contribute; $T > T_{c\sigma}$ - only diagram (a) contributes.

**Appendix B2** The $R_3$ term contributions to $3^{rd}$-order response function - diagrammatic representation for OPOP setting of incident fields.

The PP experimental set-up: $t_3 \to t$, $t_2 \to t_D$, $t_1 \to 0$, $\tau_1 = \tau_2$, $k_1 = -k_2, k_3$. Pump pulses $E_1$, $E_2$ and probe pulse $E_3$ are optical; $|E_1| = |E_2| = |E_3|$, $\omega_1 = \omega_2 = \omega_3 \equiv \omega_P$. Time ordering of incident fields in $\boldsymbol{R_3}$ term with respect to density matrix is: *bra / bra / ket*

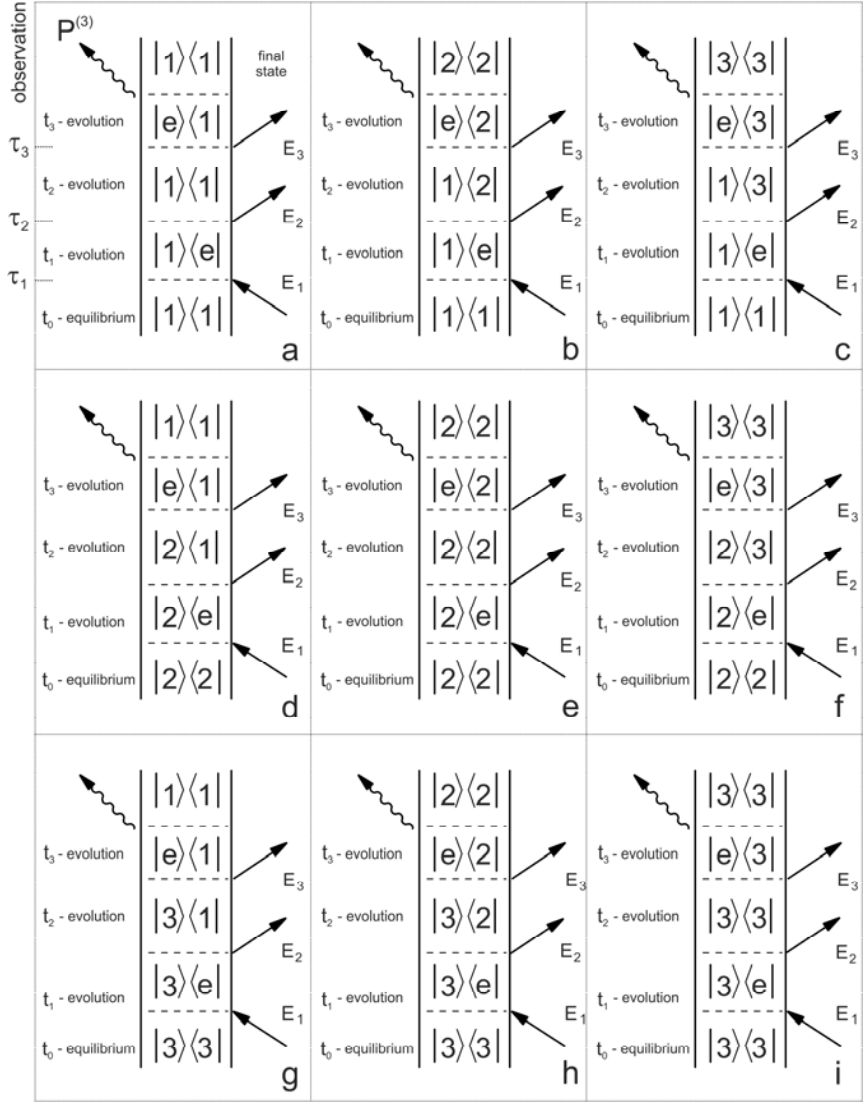

**Figure B2**  The $R_3$ term contributions to response function for OPOP spectroscopy of $MgB_2$

In temperature range:  $0 < T \leq T_{c\pi}$ - all diagrams contribute;  $T_{c\pi} < T \leq T_{c\sigma}$ - diagrams (a,c,g,i) contribute; $T > T_{c\sigma}$ - only diagram (a) contributes.

**Appendix B3** The $R_1$ term contributions to $3^{rd}$-order response function - diagrammatic representation for OPTP setting of incident fields.

The PP experimental set-up: $t_3 \to t,\ t_2 \to t_D,\ t_1 \to 0,\ \tau_1 = \tau_2,\ k_1 = -k_2,\ k_3$. Pump pulses $E_1, E_2$ are optical, $|E_1| = |E_2|$, $\omega_1 = \omega_2 \equiv \omega_P$. Probe pulse $|E_3|$, with $\omega_3 \ll \omega_P$ is from THz range. Time ordering of incident fields in $R_1$ term with respect to density matrix is: *ket / ket / ket*

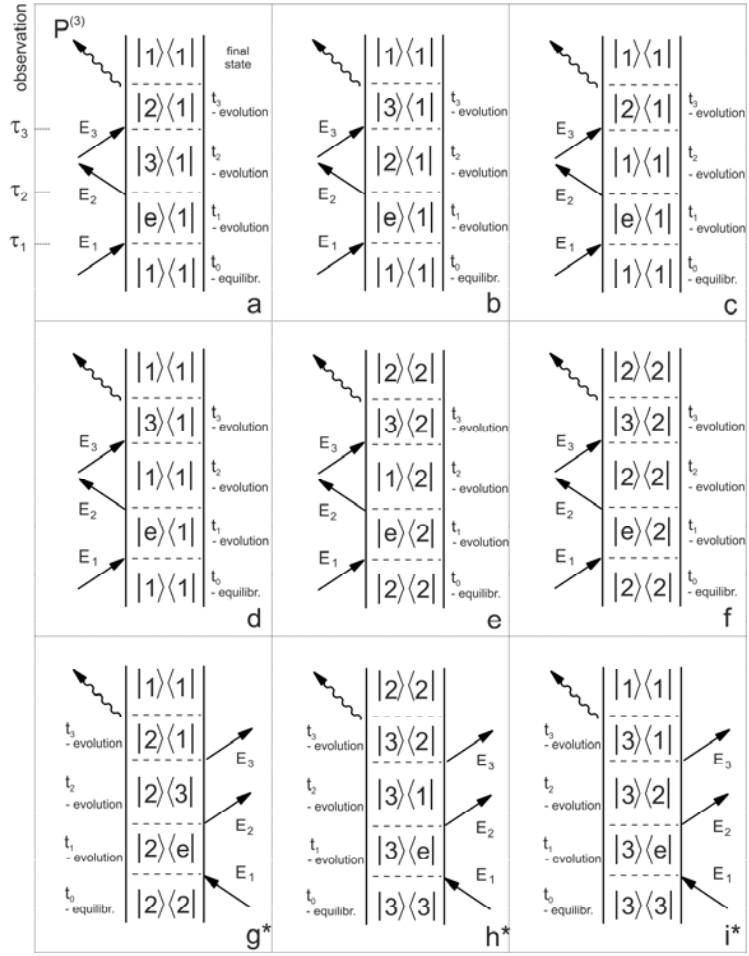

**Figure B3** The $R_1$ term contributions to response function for OPTP spectroscopy of $MgB_2$

In temperature range: $0 < T \leq T_{c\pi}$ - all diagrams contribute; $T_{c\pi} < T \leq T_{c\sigma}$ - only diagrams (d, k) contribute

**Appendix B4** The $R_3$ term contributions to $3^{rd}$-order response function - diagrammatic representation for OPTP setting of incident fields.

The PP experimental set-up: $t_3 \to t$, $t_2 \to t_D$, $t_1 \to 0$, $\tau_1 = \tau_2$, $k_1 = -k_2$, $k_3$. Pump pulses $E_1$, $E_2$ are optical, $|E_1| = |E_2|$, $\omega_1 = \omega_2 \equiv \omega_P$. Probe pulse $|E_3|$, with $\omega_3 \ll \omega_P$ is from THz range. Time ordering of incident fields in $R_3$ term with respect to density matrix is: *bra / bra / ket*

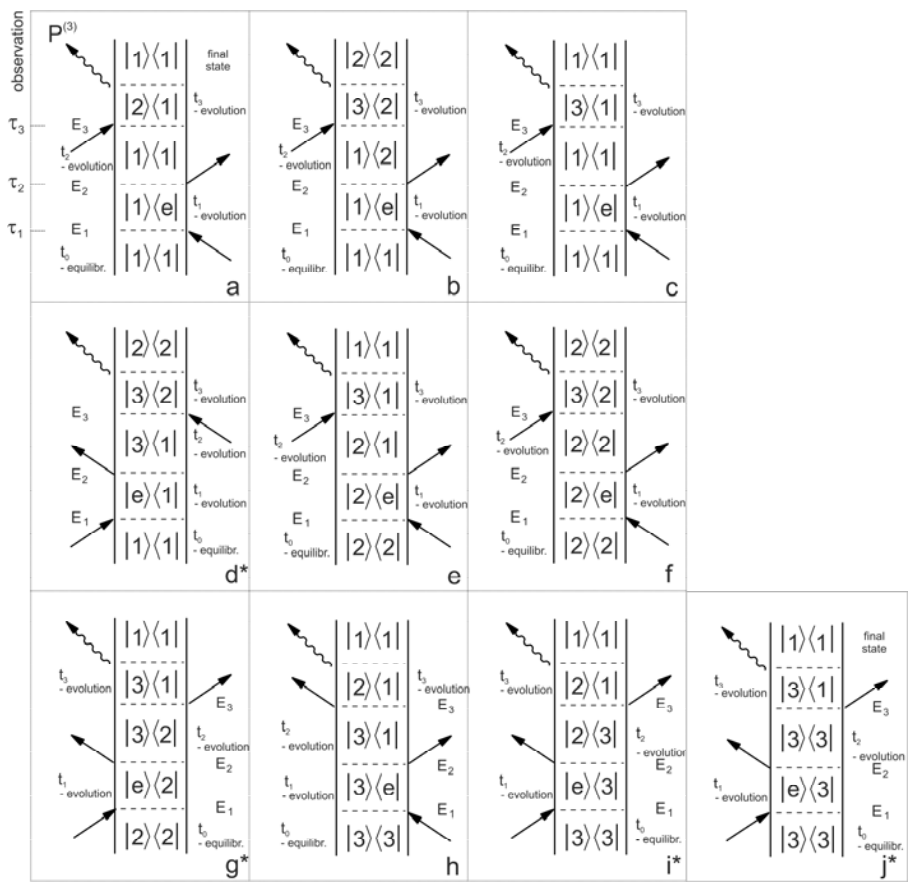

**Figure B4a** The $R_3$ term contributions to response function for OPTP spectroscopy of $MgB_2$

In temperature range: $0 < T \leq T_{c\pi}$ - all diagrams contribute; $T_{c\pi} < T \leq T_{c\sigma}$ - only diagrams (c, j) contribute and, contributions $R_1 \equiv R_3$ with final state $|1\rangle\langle 1|$.

For OPTP setting, terms $R_2$ and $R_4$ do not contribute to coherent signal. In both cases, after the first two OP pulses, the $|e\rangle\langle e|$ state is generated, and THz probe pulse is unable to stimulate electronic transition into $|g\rangle\langle g|$ states, as demonstrated by next two diagrams.

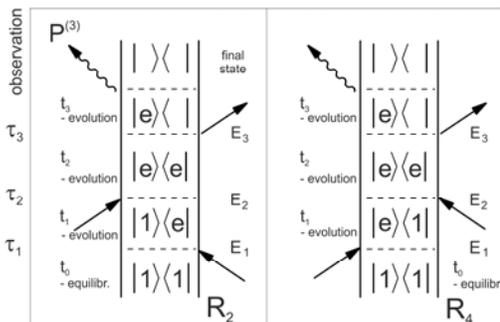

**Figure B4b** The $R_2$, $R_4$ term which do not contribute to response function for OPTP spectroscopy of $MgB_2$, but can give rise to *spontaneous relaxation* of excited state $|e\rangle\langle e|$.

## Appendix C
## Electronic correlation functions for $MgB_2$

In the frame of the rotating wave approach (RWA) the FWM processes can be described by four-point correlation functions [29] of $g \leftrightarrow e$ transition dipole moments,

$$F(\tau_1,\tau_2,\tau_3,\tau_4) = \langle d_{ge}(\tau_1)d_{eg}(\tau_2)d_{ge}(\tau_3)d_{eg}(\tau_4) \rangle \qquad (C.1)$$

where $d_{\alpha\beta}(t) = \exp(itH_\alpha)d_{\alpha\beta}\exp(-itH_\beta)$ (in units $\hbar = 1$) and the angle brackets indicate thermal average over ground-state density matrix $\rho_\beta = \exp(-\beta H_g)/Tr[\exp(-\beta H_g)]$, i.e. $\langle ... \rangle \equiv Tr[...\rho_\beta]$.

For model of $MgB_2$ in Figure 1d of the main text holds, $d_{ge} = \{d_{g_1e}, d_{g_2e}, d_{g_3e}\}$, $d_{eg} = d_{ge}^+$. Calculation of the electronic correlation functions is straightforward. Within the RWA, the energy of the lowest $|g_1\rangle$-state is reference zero, and in notation of *Mathematica* code [40] we can write,

$$H_g = \{\{0,0,0\},\{0,e_1,0\},\{0,0,e_2\}\} \qquad (C.2)$$

The quantities $e_i$ (i=1,2) mimic the energies of the superconducting gap states, i.e. $e_1 = \Delta_\pi(T)$ and $e_2 = \Delta_\sigma(T)$ The energy of the $|e\rangle$-state is $\omega_{ge} = \omega_P = 1.5eV$ and, since in the model there is only single excited state energy level $\varepsilon_e$, it can serve as the excited state reference zero, i.e. $H_e = 0$.

Due to time-translation invariance and Hermitean properties of the involved operators, only four of the correlation functions (C.1) are independent-see A14a-d. For PP description only two of them (A14b,c) are relevant, i.e.

$$R_2(t_3,t_2,t_1) = F(0,t_1+t_2,t_1+t_2+t_3,t_1) \qquad (C.3)$$

and

$$R_3(t_3,t_2,t_1) = F(0,t_1,t_1+t_2+t_3,t_1+t_2) . \qquad (C.4)$$

In PP set-up, the first-two interactions with external pulse (laser) take place at the same time (time-contracted pulses, $t_1 \to 0$, $k_1 = -k_2$). The pump-probe delay is denoted as $t_D$ (i.e. $t_2 \to t_D$) and $t$ stands for time after the probe-pulse (i.e. $t_3 \to t$).

Using the symbolic software of *Mathematica* [40], a straightforward application of methods presented here leads to the desired electronic correlation functions used in Equation (12) of the main text, namely, the terms $R_i(t,t_D) \equiv R_i(t,t_D,0)$ (i=2,3). For these terms has been derived,

$$R_2(t,t_D) = \langle d_{ge}d_{eg} \exp(iH_g t) d_{ge}d_{eg} \rangle$$
$$R_3(t,t_D) = \langle d_{ge}d_{eg} \exp(iH_g(t+t_D)) d_{ge}d_{eg} \exp(-iH_g t_D) \rangle$$

The explicit temperature-dependent form of *R*-terms corresponding to the model displayed in Figure 1d, with notation of (C2) is,

$$R_2(t,t_D) = \frac{\left(|d_{g_1e}|^2 + |d_{g_2e}|^2 e^{ite_1} + |d_{g_3e}|^2 e^{ite_2}\right)\left(|d_{g_1e}|^2 + |d_{g_2e}|^2 e^{-\beta e_1} + |d_{g_3e}|^2 e^{-\beta e_2}\right)}{1 + e^{-\beta e_1} + e^{-\beta e_2}}$$

and                                                                                                                              (C.5)

$$R_3(t,t_D) = \frac{\left(|d_{g_1e}|^2 + |d_{g_2e}|^2 e^{i(t+t_D)e_1} + |d_{g_3e}|^2 e^{i(t+t_D)e_2}\right)\left(|d_{g_1e}|^2 + |d_{g_2e}|^2 e^{-it_D e_1 - \beta e_1} + |d_{g_3e}|^2 e^{-it_D e_2 - \beta e_2}\right)}{1 + e^{-\beta e_1} + e^{-\beta e_2}}$$

The function $R_2$, which represents SE process, depends only on time $t$ (i.e. time after the probe-pulse), whereas the function $R_3$ contributing to GSB depends on both times- $t$ and $t_D$. Thus, from the point of view of time-integrated signal, Equation (11), the SE part behaves like a renormalized two-level system.

As compared to the standard molecular models of nonlinear optics, with a single-level ground and multi-level excited state, the time-dependence of correlation functions $R_2$ and $R_3$ is completely opposite because in molecular systems the SE part, through PP-time dependence on $t_D$, bears information on the excited state spectrum.

A possibility for an introduction of some general-global dephasing processes, which can not be described by a line-shape function $g(t)$ - (Appendix D), deserves to be mentioned. If we include a general relaxation of populations (diagonal part of density matrix) and coherences (non-diagonal, $g-e$ part of density matrix), by so-called Bloch model [29] represented by relaxation times $T_1$ (population relaxation) and $T_2$ (dephasing relaxation), then the electronic correlation function can be written as,

$$R_2(t_3,t_2,t_1) \rightarrow e^{-(t_3+t_1)/T_2} e^{-t_2/T_1} R_2(t_3,t_2,t_1)$$
$$R_3(t_3,t_2,t_1) \rightarrow e^{-(t_3+t_1)/T_2} R_3(t_3,t_2,t_1)$$

Thus, our PP correlation functions, (C.5), are replaced by,

$$R_2(t,t_D) \rightarrow e^{-t/T_2} e^{-t_D/T_1} R_2(t,t_D)$$ (C.5a)
$$R_3(t,t_D) \rightarrow e^{-t/T_2} R_3(t,t_D)$$

For relaxation times $T_1$ and $T_2$ must hold,

$$1/T_2 \geq 1/(2T_1)$$

This condition bounds the dephasing relaxation: $T_2 \leq 2T_1$. The net result is that the Bloch relaxation results in a common, short-lived, decay on both electronic correlation functions and a long-lived population decay on $R_2$ function. Due to fact that our aim is to analyze the influence of temperature-dependent bath on our electronic system, we neglected the $T_2$ contribution and preserve the last, long-lived $T_1$ decay. Presence of the only $T_1$ decay term over time period $t_D \equiv t_2$ in $R_2$ term follows directly from the diagrams in Figure B1, whilst to decay process covered by $R_3$ term over the time period $t_D \equiv t_2$ contributes mainly coherences - dephasing relaxation, see Figure B2.

**Appendix D**
**Vibration correlation functions derived from Eliashberg function of MgB$_2$**

The influence of the dissipative environment is calculated within the multimode Brownian oscillator model [29], when the local vibrations act as a collective bath on the g - e electronic system. In the frame of cumulant expansion (A25), the corresponding correlation functions are [30],

$$R_2^v(t_3,t_2,t_1) = \exp\left(-g^*(t_3) - g^*(t_1) + f_+^*(t_3,t_2,t_1)\right)$$ (D.1)

$$R_3^\nu(t_3,t_2,t_1) = \exp\left(-g(t_3) - g^*(t_1) + f_-^*(t_3,t_2,t_1)\right)$$

The superscript $\nu$ is used to distinguish vibration from electronic functions and, the auxiliary functions are introduced,

$$f_+(t_3,t_2,t_1) = g^*(t_2) - g^*(t_2+t_3) - g(t_1+t_2) + g(t_1+t_2+t_3) \quad \text{(D.2)}$$
$$f_-(t_3,t_2,t_1) = g(t_2) - g(t_2+t_3) - g(t_1+t_2) + g(t_1+t_2+t_3)$$

whilst the asterisk stands for complex conjugate. The $g(t)$ is the complex line-shape function [29,30], $g(t) = g_R(t) + ig_I(t)$, with positive real part.

The $g(t)$ underlies time-symmetry, $g_R(-t) = g_R(t)$, $g_I(-t) = -g_I(t)$.

For PP setting ($t_3 \to t$, $t_2 \to t_D$, $t_1 \to 0$) we get,

$$R_2^\nu(t,t_D) \equiv R_2^\nu(t,t_D,0) = \exp[-g^*(t) + 2i(g_I(t_D) - g_I(t+t_D))] \quad \text{(D.3)}$$
$$R_3^\nu(t,t_D) \equiv R_3^\nu(t,t_D,0) = \exp[-g(t)]$$

The particular form of the line-shape function depends on the specified dissipative model, which is defined through correlation function $C(\tau)$, i.e. (in units $\hbar = 1$)

$$g(t) = \int_0^t d\tau \int_0^\tau d\tau_1 C(\tau_1) \quad \text{(D.4a)}$$

Correlation function is generally related to the spectral density $J(\omega)$,

$$C(t) = \frac{1}{\pi} \int_{-\infty}^{\infty} d\omega\, J(\omega) \frac{e^{-i\omega t}}{1 - e^{-\beta\omega}}, \quad \text{(D.4b)}$$

where $\beta$ is the inverse temperature. In present study, we decompose the spectral density in Meier-Tannor form [26,27] of Lorentzians,

$$J(\omega) = \sum_{k=1}^{n} p_k \frac{\omega}{[(\omega+\Omega_k)^2 + \Gamma_k^2][(\omega-\Omega_k)^2 + \Gamma_k^2]} \quad \text{(D.4c)}$$

The number of components (n), appropriate fit parameters ($p_k$, $\Omega_k$, $\Gamma_k$) have been chosen for each type of coupling ($\lambda_\pi$, $\lambda_{\sigma\pi}$, $\lambda_\sigma$, $\lambda_{\pi\sigma}$). We note that spectral density $J(\omega)$ is linearly proportional to the dimensionless $\alpha^2 F(\omega)$ function, i.e. $J(\omega) = K\omega\,\alpha^2 F(\omega)$, where $K$ is a dimensionless prefactor. According to [19], we have used the normalized form of $\alpha^2 F(\omega)$

$$\lambda = 2\int_0^\infty \alpha^2 F(\omega) \frac{d\omega}{\omega} \quad \text{(D.5)}$$

and this form has been used for Meier-Tannor decomposition of $J(\omega)$, equation (D.4c).

Finally, for particular temperature range, the corresponding coupling types obey the coupling scheme of AAT or BCS model, as specified in Table 1 of the main text.

As regards calculation of line-shape function in wide range of temperatures (5K - 300 K), rather than application of the standard Matsubara decomposition [27], we implemented the [N−1/N] Padé spectrum decomposition [28] of the Bose function $1/(1 - e^{-\beta\omega})$,

$$\frac{1}{1 - e^{-x}} = \frac{1}{x} + \frac{1}{2} + \sum_{j=1}^{N} \left(\frac{\eta_j}{x + i\xi_j} + \frac{\eta_j}{x - i\xi_j}\right), \quad \text{(D6)}$$

where the expansion coefficients $\eta_j$ and roots $\xi_j$, for a selected N, have to be calculated. Due to the fact that the actual frequencies of the spectral density function, $\Omega_k$, are in the range 250 cm$^{-1}$ - 730 cm$^{-1}$ and temperatures are from interval 5K - 300 K, the Padé decomposition with N=64 seems to be satisfactory within relative accuracy $1.10^{-3}$ % ($\beta\Omega_k$ is in the range 10 - 2500).

By means of the residue theorem of complex analysis implemented in Mathematica$^©$ code [40], the infinite integrals in (D4b) for $t \geq 0$ have been symbolically calculated. The results have been used for numerical calculation.

**Appendix E**
**Form of vibration correlation function for a single-mode/single-component Tannor model**

Let as analyze a single coupling-type (single-mode), which is represented by only a single Lorenzian component in its spectral density (or in Eliashberg) function.
The Tannor form for such a single-mode/ single-component spectral density is,

$$J(\omega) = \omega \frac{p}{[(\omega+\Omega)^2 + \Gamma^2][(\omega-\Omega)^2 + \Gamma^2]} \tag{E1}$$

with a dimension of frequency. It is characterized by a central frequency $\Omega$, by a width $\Gamma$ and a weight $p$ (with dimension, $\dim(\omega)^4$).

The spectral density can be characterized by frequency,

$$\Omega_J = \frac{1}{\pi}\int_0^\infty \frac{J(\omega)}{\omega} d\omega \tag{E2}$$

This frequency is rather known from a literature under the term, *reorganization frequency (energy)*.

For a single-mode/ single-component form of $J(\omega)$ - (E1) we get,

$$\Omega_J = \frac{p}{4\Gamma(\Omega^2 + \Gamma^2)} \tag{E3}$$

For multi-component Tannor model the reorganization energy is simply a sum of particular Lorenzian- contributions of particular type of coupling.

For parameters used in our simulation, we obtain (after summing-up the proper number of components for each type of coupling, e.g for $\sigma-\sigma$ contribution two contributions, for $\pi-\pi$ and $\sigma-\pi$ four contributions - see Figures 2a, b in the main text) the following results –Table E1,

| Interaction mode – type of coupling | $\Omega_J \ [cm^{-1}]/[meV]$ | $\lambda$ |
|---|---|---|
| $\sigma-\sigma$ | 93.0 / 11.5 | 1.017 |
| $\pi-\pi$ | 35.8 / 4.4 | 0.448 |
| $\sigma-\pi$ | 14.0 / 1.74 | 0.213 |

**Table E1** Reorganization energy pertaining to particular type of coupling

The first column designates the type of coupling, in the second column are calculated reorganization energies and in the last column are published [19] dimensionless EP coupling constants calculated as an integral norm according to,

$$\lambda = 2\int_0^\infty \alpha^2 F(\omega) \frac{d\omega}{\omega} \quad (E4)$$

The correlation function,

$$C(t) = \frac{1}{\pi} \int_{-\infty}^\infty d\omega\, J(\omega) \frac{e^{-i\omega t}}{1 - e^{-\beta\hbar\omega}} \quad (E5)$$

with inverse temperature $\beta$, can be calculated by method of residues of complex function [40].

The spectral function (E.1) has poles in the complex plane at four frequencies, $\pm\Omega \pm i\Gamma$. The Bose function, $1/(1-e^{-x})$, can be approximated by N-fold Padé decomposition [28]. We divide it into two parts,

$$f_N^{Bose}(x) = \frac{1}{x} + \frac{1}{2} + \sum_{j=1}^N \left( \frac{\eta_j}{x + i\xi_j} + \frac{\eta_j}{x - i\xi_j} \right) = \\ = f_0^{Bose}(x) + f_{NR}^{Bose}(x) \quad (E6)$$

where $f_0^{Bose}(x) = 1/x + 1/2$ means exactly the series expansion of $1/(1-\exp(-x))$ around $x=0$ (i.e. part of high-temperature limit). The second part of the decomposition, $f_{NR}^{Bose}$, has poles at points $\mp i\xi_j$. For times $t>0$ in equation (E.2) we use contour integration in the lower half plane of the complex energy plane (the contour integral closing over the lower semi-infinite circle, due to the presence of $\exp[-i\omega t]$ part, is zero). We use residues [40] of argument at complex poles at $\pm\Omega \pm i\Gamma$ and at $-i\xi_j/\beta\hbar$. To have an overview about the values of Padé decomposition coefficients, Figure E1 shows the position of $\{\xi_j/(2\pi j),\ \eta_j\}_{j=1}^{64}$ pairs for the order N=64 (this has been used in the main text).

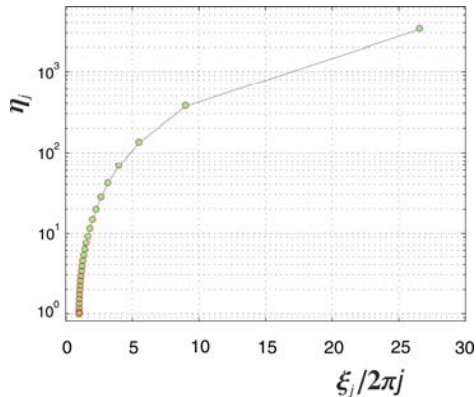

**Figure E1** Padé decomposition of Bose function for N=64 (lin-log scale).

As it has been mentioned in the main text, we have used high N-order decomposition in order to achieve relative accuracy better then $10^{-5}$ for temperature region 5K - 300 K and for frequencies (energies) in the range of 250 cm$^{-1}$ - 730 cm$^{-1}$. The product of the

Tannor-disassembled spectral density and that of Padé decomposed Bose function, yields a sum of simple inverse terms of $1/(\omega-\Psi_i)$ form. The complex poles at $\Psi_i$ in the lower frequency plane result in Eq.(E.5) by a straightforward use of the residue theorem [40].

The correlation function can be divided according to division of the Bose function in equation (E.6). The final form reads

$$C(t) = C_{1,R}(t) - iC_{1,I}(t) + C_{2,R} \tag{E7}$$

where $C_{1,R}(t) - iC_{1,I}(t)$ corresponds to $f_0^{Bose}$ and $C_{2,R}$ to $f_{NR}^{Bose}$. The analytic form of these parts is;

$$C_{1,R}(t) = \frac{p(\Gamma \sin(t\Omega) + \Omega \cos(t\Omega))}{2\beta\hbar\,\Gamma\Omega(\Gamma^2+\Omega^2)} e^{-\Gamma t} \tag{E8}$$

for the real part and, corresponding imaginary part reads,

$$C_{1,I}(t) = \frac{p\sin(t\Omega)}{4\Gamma\Omega} e^{-\Gamma t} \tag{E9}$$

Finally, the $f_{NR}^{Bose}$ contribution is a real function,

$$C_{2,R}(t) = \frac{p}{\beta\hbar} \sum_{j=1}^{N} \frac{\eta_j}{\left((\Gamma-\nu_j)^2+\Omega^2\right)\left((\Gamma+\nu_j)^2+\Omega^2\right)} \tag{E.10}$$

$$\left[\frac{\Gamma(\Gamma^2-\nu_j^2+\Omega^2)\sin(t\Omega)+\Omega(\Gamma^2+\nu_j^2+\Omega^2)\cos(t\Omega)}{\Gamma\Omega} e^{-\Gamma t} - 2\nu_j e^{-t\nu_j}\right]$$

where the Matsubara-like frequencies are introduced $\nu_j = \xi_j/(\beta\hbar)$

Following the above mentioned results, we can accordingly decompose the (dimensionless) line-shape function, too.

$$g(t) = \int_0^t d\tau \int_0^\tau d\tau_1 C(\tau_1) \tag{E11}$$

We have,

$$g(t) = g_{1,R}(t) - ig_{1,I}(t) + g_{2,R}(t) \tag{E12}$$

Explicit form of particular terms in (E12) is,

$$g_{1,R}(t) = \frac{t}{\tau}$$
$$- \frac{p\left(\Omega(\Omega^2-3\Gamma^2)\cos(t\Omega) - \Gamma(\Gamma^2-3\Omega^2)\sin(t\Omega)\right)}{2\beta\hbar\Gamma\Omega(\Gamma^2+\Omega^2)^3} e^{-\Gamma t} + \frac{p(\Omega^2-3\Gamma^2)}{2\beta\hbar\Gamma(\Gamma^2+\Omega^2)^3} \tag{E13}$$

and

$$g_{1,I}(t) = \Omega_J t$$
$$+ \frac{p\left(2\Gamma\Omega\cos(t\Omega) - (\Omega^2-\Gamma^2)\sin(t\Omega)\right)}{4\Gamma\Omega(\Gamma^2+\Omega^2)^2} e^{-\Gamma t} - \frac{p}{2(\Gamma^2+\Omega^2)^2} \tag{E14}$$

The frequency $\Omega_J$ is defined by equation (E3) and, in (E13) we have introduced the temperature-dependent decay time,

$$\tau = \beta\hbar(\Gamma^2+\Omega^2)^2 / p \tag{E15}$$

which governs the long-time behavior of the real part, $g_{1,R}(t)$. As it can be expected, the imaginary part, $g_{1,I}(t)$, is temperature-independent.

Finally, the form of contribution originating from term $f_{NR}^{Bose}$, i.e. $g_{2,R}(t)$, is a little-bit cumbersome,

$$g_{2,R}(t) = \frac{p}{\beta\hbar}\sum_{j=1}^{N}\left\{\frac{\eta_j(2\Gamma+\nu_j)}{\nu_j\Gamma(\Gamma^2+\Omega^2)((\Gamma+\nu_j)^2+\Omega^2)}\right.$$
$$-\frac{2\eta_j}{\nu_j\left(\Gamma^4-2\Gamma^2(\nu_j^2-\Omega^2)+(\nu_j^2+\Omega^2)^2\right)}e^{-t\nu_j}$$
$$\left.+\frac{\eta_j\left(\Gamma\sin(t\Omega)(\Gamma^2-\nu_j^2-3\Omega^2)-\Omega\cos(t\Omega)(-3\Gamma^2+\nu_j^2+\Omega^2)\right)}{\Gamma\Omega(\Gamma^2+\Omega^2)\left(\Gamma^4-2\Gamma^2(\nu_j^2-\Omega^2)+(\nu_j^2+\Omega^2)^2\right)}e^{-\Gamma t}\right\} \quad (E16)$$

As it can be seen from (E16), the $f_{NR}^{Bose}$ part contributes to the real part of $g(t)$.

It is important to note some facts that follow from numerical simulation;
As it can be seen, the first part of $g_{1,R}(t)$ behaves as $g_{1,R}(t) \approx t/\tau$, whilst decay time $\tau$ decreases with increasing temperature, but substantially grows with the width of peak and corresponding frequency (see equation (E.15)). It means that relaxation dynamics ($\propto \exp(-g(t))$) is faster with rising temperature, but relaxation process is slow-down by increase of the width and corresponding frequency of the peak in particular EP coupling mode. This decay term is slightly modulated, however, by contribution of $g_{2,R}(t)$ - equation (E16). Namely, in the low-temperature limit ($1/\beta \to 0$), expansion to second order in $1/\beta$ yields a contribution, and the decay time is modified,

$$\tau = \beta\hbar(\Gamma^2+\Omega^2)^2 / \left[p(1+2\sum_{j=1}^{N}\eta_j)\right] \quad (E17)$$

As regards the imaginary part of line-shape function, $g_{1,I}(t)$, it constantly decreases with parameter $\Omega_J$ (see equation (E.3)). Moreover, its time shape is modulated by decaying term proportional to $e^{-\Gamma t}$. Thus, for times $t > 1/\Gamma$ only the linear term persists.

Finally, we note the form of so-called spectral-diffusion part [29] of the PP signal, defined as $f(t,t_D) = -2i(\text{Im}[g(t_D)] - \text{Im}[g(t+t_D)])$.

Series expansion in time $t$ has the form,

$$f(t,t_D) = -2i\Omega_J t\left(1 - e^{-\Gamma t_D}\cos(t_D\Omega) - e^{-\Gamma t_D}\frac{\Gamma\sin(t_D\Omega)}{\Omega}\right) \quad (E18)$$

i.e. this term is unimportant for PP times $t_D > 1/\Gamma$.

A final note: for Bose function in the zero temperature limit we have,

$$1/(1-\exp(-\beta\hbar\omega)) \xrightarrow{\beta \to \infty} H(\omega)$$

where $H(\omega)$ is the Heaviside step function (0 for $\omega < 0$ and 1 for $\omega \geq 0$). Then, the formal expression for line-shape function is:

$$g_0(t) = \frac{1}{\pi} \int_0^\infty \frac{J(\omega)}{\omega^2} \left(1 - e^{-it\omega}\right) d\omega - it\Omega_J \qquad \text{(E19)}$$

It means that in zero-temperature limit, an analytical form of line-shape function $g_0(t)$ for single-mode/single-component Tannor spectral model - Eq.(E1) does not exist.

A remark to δ-peak frequency spectral model:

A single-component spectral "density" represented by a δ-peak at frequency $\omega_0$ and reorganization frequency $\Omega_J$ (defined in equation (E.2) ) equals to,

$$J(\omega) = \pi \, \Omega_J \, \omega_0 \left[\delta(\omega - \omega_0) - \delta(\omega + \omega_0)\right] \qquad \text{(E20)}$$

with zero spectral width .

The correlation function can be calculated according to equation (E5) and we get the line-shape function $g(t)$ in a well-known text-book form,

$$g(t) = \frac{\Omega_J}{\omega_0}\left(1 - \cos(t\omega_0)\right)\coth\left(\frac{\beta\hbar\omega_0}{2}\right) \\ - i\left(t\Omega_J - \frac{\Omega_J}{\omega_0}\sin(t\omega_0)\right) \qquad \text{(E21)}$$

The real part is a periodic function of time and imaginary part decays (or is minus proportional to) with reorganization frequency $\Omega_J$. In fact, the ratio $\Omega_J/\omega_0$ corresponds to the so-called Huang-Rhys factor,

$$X = \frac{1}{\pi}\int_0^\infty \frac{J(\omega)}{\omega^2} d\omega \ , \qquad \text{(E22)}$$

which for Tannor spectral density (E1) diverges.

The Huang-Rhys factor is a dimensionless coupling constant, which should be compared with $\lambda$ if this parameter is calculated from the second moment of Eliashberg function at some "relevant" frequency $\omega_0$. However, as it can be seen from Figures 2a,b in the main text, Eliashberg function for each coupling type is represented by a multicomponent Lorenzian and, in this circumstances, it is hard to decide which is "relevant" frequency for superconductivity- see Discussion in the main text.